\documentclass[a4paper,11pt]{article}
\usepackage{caption}
\pdfoutput=1 % if your are submitting a pdflatex (i.e. if you have
\usepackage{jheppub} % for details on the use of the package, please
\usepackage[T1]{fontenc} % if needed
\usepackage{tabularx}
\usepackage{xcolor}
\usepackage{booktabs} 
\usepackage{subcaption}

\usepackage{float}

\allowdisplaybreaks

\title{\boldmath Asymptotic expansion of cavity Quasi-Normal Modes in Schwarzschild de-Sitter black hole}

\author{Arnab Priya Saha,}
\author{Prajwal Shivanna}

\affiliation{Centre for High Energy Physics, Indian Institute of Science,\\ C. V. Raman Avenue, Mathikhere, Bengaluru 560012, India}

\emailAdd{arnabsaha@iisc.ac.in}
\emailAdd{prajwals1@iisc.ac.in}

\abstract{We study large-overtone  quasi-normal modes   of a massive scalar field in a four-dimensional  Schwarzschild de-Sitter black hole in the presence of a wall placed at the static sphere. Dirichlet boundary condition is imposed at the wall. At leading order the quasi-normal modes with large overtones scale as $n$ for $n\gg 1$ and are inversely proportional to the bouncing time. We analyze correction terms up to $n^{-1}$ order in the asymptotic expansion. }

\begin{document} 
\maketitle
\flushbottom

\section{Introduction}
In black hole perturbation theory, Quasi-Normal Modes (QNM) at large overtones play an important role in understanding the interior geometry of black holes. See \cite{Berti:2009kk, Konoplya:2011qq} for a review of QNM in black hole physics. Large quasi-normal mode frequencies have been studied for a wide class of black holes in \cite{ Leaver1985, PhysRevD.47.5253, Hod:1998vk, Motl:2002hd, Cardoso:2003sw, Motl:2003cd, Berti:2003jh, Cardoso:2003vt, Andersson:2003fh, Choudhury:2003wd, Berti:2004um, Cardoso:2004up, Natario:2004jd, Das:2004db, Keshet:2007be, Festuccia:2008zx, Oshita:2026vxh} using various methods such as continued fraction, numerical investigation, monodromy matching and WKB analysis.

The interior of black holes can be probed by certain class of complex geodesics, known as bouncing geodesics \cite{Fidkowski:2003nf, Festuccia:2008zx, Festuccia:2005pi, Faruk:2023uzs, Aalsma:2026kqj}. These geodesics are null limit of timelike or spacelike geodesics at high energy which bounce off from the near-singularity region of black hole interior \cite{Grozdanov:2026cut}. In Schwarzschild black hole, convergence region of QNM expansion has been shown to be characterized by bouncing geodesic \cite{Arnaudo:2026tcy}. Recently, the relation between asymptotic QNM and bouncing time in AdS black holes with Kasner-type singularities has been investigated in \cite{Hartnoll:2026vhu,Xiao:2026pir}. From a holographic perspective, bouncing geodesics are important for their relation to boundary correlators \cite{PhysRevD.103.066018, Grinberg:2020fdj, Giombi:2026kdz, Grozdanov:2026cut,Afkhami-Jeddi:2025wra}.

Here we consider a probe scalar field in a Schwarzschild black hole in de-Sitter space (SdS). A time-like surface, acting as a reflecting wall \cite{Grozdanov:2026lnc},  is placed at a radial distance coinciding with the static sphere. The scalar field wave function satisfies ingoing boundary condition at the black hole horizon. We impose Dirichlet boundary condition at the  wall, implying that wave function vanishes at the static sphere radius. Although flux of the scalar wave function vanishes at the surface, non-zero solution exists beyond the wall, resulting in having both ingoing and outgoing plane waves at the cosmological horizon.  We refer to the frequencies of the scalar wave function, which satisfy the above mentioned boundary conditions, as cavity QNM. In this paper we analyze the asymptotic behaviour of such QNM at large overtones. 

Recent works on the analysis of QNM in SdS include \cite{Wu:2025wbp,Wu:2026hvf}. We follow WKB analysis of matching solutions across Stokes lines, similar to the methods adopted in \cite{Cardoso:2004up,Ghosh:2005aq,Hartnoll:2026vhu}. 

Our motivation behind considering the above mentioned boundary conditions follows from the analysis of \cite{Grozdanov:2026ktq}. In this reference, retarded propagator and two-sided Wightman correlator have been studied in SdS cavity set up.  In that paper, the authors have considered a time-like surface satisfying  Dirichlet, Neumann or Robin boundary conditions. There the domain of the physical solution is defined between the black hole horizon and the time-like wall. Importantly, boundary correlator for a putative theory living on the time-like surface has been constructed from the Wightman correlator. The boundary correlator obeys a product formula of the form \cite{Dodelson:2023vrw}
\begin{equation}\label{bound-cor}
	G_{12}^{\partial}\left(\omega\right) = \frac{G_{12}^{\partial}\left(0\right)}{\prod_{n=1}^{\infty}\left(1-\frac{\omega^{2}}{\omega_{n}^{2}}\right)\left(1-\frac{\omega^{2}}{\left(\omega_{n}^{\ast}\right)^{2}}\right)}~.
\end{equation}
Poles of the boundary correlator are  $\omega=\pm\omega_{n}, \; \pm\omega^{\ast}_{n}$. $\omega_{n}$ are the zeroes of Wronskian of the retarded Green's function, which is obtained from two Jost solutions, $g\left(\omega,z\right)$ and $h\left(\omega,z\right)$. These functions have the following boundary conditions: $g\left(\omega,z\right)$ vanishes at the boundary wall and $h\left(\omega,z\right)$ has ingoing boundary condition at the black hole horizon. Wronskian of the two Jost functions becomes proportional to $h\left(\omega,z\right)$ evaluated at the wall. Therefore,  $h\left(\omega,z\right)=0$ at the boundary wall corresponds to QNM.  By studying the properties of the boundary correlator at high energy, leading asymptotic expression for large QNM in terms of the bouncing time has been derived in that paper. This relation has the form $\omega_{n}t_{\ast}=2\pi n$, for large integer values of $n$.

We have worked out an asymptotic expansion of large QNM to $\frac{1}{n}$ order. The result can be summarized as $\omega_{n} z_{0}= \pi n+A_{0} + \frac{A_{-\frac{1}{2}}}{\sqrt{n}}+\frac{A_{-1}}{n} + \mathcal{O}\left(n^{-\frac{3}{2}}\right)$, with $2z_{0}$ being the bouncing time corresponding  to the static sphere.  We can draw an analogy with the cavity wall set up described in \cite{Grozdanov:2026ktq} by treating the region between black hole horizon and the time-like surface with Dirichlet boundary condition as the physical part of the spacetime and  wave function in region beyond the surface to be analytic continuation of solution to the scalar equation of motion. 

This paper is organized as follows: a brief review of Schwarzschild dS black hole and bouncing geodesic in it is presented in Sec.(\ref{Sec:SdS}). Then we discuss asymptotic QNM in SdS cavity set up and give the derivation to  sub-leading order expansion of the large overtones in Sec.(\ref{Sec:scalar}). In Sec.(\ref{Sec:subleading}) we derive the corrections to the asymptotic expansion beyond sub-leading order. We perform numerical checks of our derived expressions in Sec.(\ref{Sec:Num}). We conclude with a discussion of our results in Sec.(\ref{Sec:discuss}). Appendix (\ref{App:Stokes}) contains details of Stokes topology. Some details of the calculation leading to sub-leading order corrections of asymptotic QNM are given in Appendix (\ref{App:Volterra}). 

\section{Schwarzschild dS black hole}
\label{Sec:SdS}
We briefly review the $\left(3+1\right)$-dimensional geometry of a Schwarzschild black hole in de-Sitter space. The metric is given by 
\begin{equation}\label{SdS-metric}
	\mathrm{d}s^{2} = -f\left(r\right)\mathrm{d}t^{2}+\frac{1}{f\left(r\right)}\mathrm{d}r^{2}+ r^{2}\;\mathrm{d}\Omega_{2}^{2}~,
\end{equation}
where $t$ and $r$ are the time and radial coordinates respectively, and the two-dimensional sphere metric is $\mathrm{d}\Omega_{2}^{2}=\mathrm{d}\theta^{2}+\sin^{2}\theta\; \mathrm{d}\phi^{2}$ with $0\le\theta\le\pi$ and $0\le\phi<2\pi$. $f\left(r\right)$ depends on mass of the black hole, $M$ and cosmological constant, $\Lambda$ and has the form
\begin{equation}
	f\left(r\right)=1-\frac{2GM}{r}-\frac{\Lambda}{3}r^{2}~.
\end{equation}
There are three roots of $f\left(r\right)$. Black hole horizon $r_{b}$ and cosmological horizon $r_{c}$ correspond to the two positive roots with $r_{c}>r_{b}>0$. The third root, $r_{f}$ satisfies $r_{f}=-\left(r_{b}+r_{c}\right)$ and hence is negative. We can also write the following equations,
\begin{equation}
	r_{b}^{2}+r_{b}r_{c}+r_{c}^{2}=\frac{3}{\Lambda}~, \qquad r_{b}r_{c}\left(r_{b}+r_{c}\right)=\frac{6GM}{\Lambda}~.
\end{equation}
Above equations have the solutions,
\begin{equation}
	r_{b}=\frac{2}{\sqrt{\Lambda}}\cos\left(\theta+\frac{\pi}{3}\right)~, \quad r_{c}=\frac{2}{\sqrt{\Lambda}}\cos\left(\theta-\frac{\pi}{3}\right)~, \qquad \theta =\frac{1}{3} \cos^{-1}\left(3GM\sqrt{\Lambda}\right)~.
\end{equation}
In the limit, $M\rightarrow 0$ we recover the pure static patch, which has $r_{c}=\sqrt{\frac{3}{\Lambda}}$ and the black hole horizon moves to $r=0$.

It is convenient for later purpose to use tortoise coordinate $z$, defined as
\begin{equation}
	\frac{\mathrm{d}z}{\mathrm{d}r} = -\frac{1}{f\left(r\right)}~.
\end{equation}
We choose $z=0$ at the static sphere radius, $r_{\mathcal{O}}=\left(\frac{3GM}{\Lambda}\right)^{\frac{1}{3}}$. A static observer at $r_{\mathcal{O}}$ has zero proper acceleration, so we have the condition: $f'\left(r_{\mathcal{O}}\right)=0$. For $3GM\sqrt{\Lambda}<1$, static sphere lies between the two horizons, $r_{b}<r_{\mathcal{O}}<r_{c}$ and we will restrict to this regime. In the region, $r_{b}<r<r_{c}$  tortoise coordinate can then be given by 
\begin{equation}\label{closed-tortoise}
	z\left(r\right)=-\sum_{i\in\{b,c,f\}}\frac{1}{f'\left(r_{i}\right)}\log\left\vert\frac{r-r_{i}}{r_{\mathcal{O}}-r_{i}}\right\vert~.
\end{equation}
If $r<r_{b}$, then the contour of integration passes through $f\left(r_{b}\right)=0$. Deforming the integration contour on the lower half $r$-plane gives an imaginary shift\footnote{We have specifically chosen this contour to be consistent with the negative sign for the imaginary part of the quasi-normal modes. Here $\operatorname{Im}\left(z_{0}\right)=\frac{\beta_{b}}{4}$, which is related to temperature of the black hole. },
\begin{equation}\label{contour-prescription}
	z\left(r\right)=-\sum_{i\in\{b,c,f\}}\frac{1}{f'\left(r_{i}\right)}\log\left\vert\frac{r-r_{i}}{r_{\mathcal{O}}-r_{i}}\right\vert +\frac{i\pi}{f'\left(r_{b}\right)}~, \qquad r<r_{b}~.
\end{equation}
Since $f'\left(r_{b,c}\right)=\pm \frac{\Lambda}{3r_{b,c}}\left(r_{c}-r_{b}\right)\left(r_{b,c}-r_{f}\right)$, therefore from Eq.\eqref{closed-tortoise} we can see that $z\rightarrow\infty$ as $r\rightarrow r_{b}$ and $z\rightarrow-\infty$ as $r\rightarrow r_{c}$.

\subsection{Bouncing geodesic}
Since the metric in Eq.\eqref{SdS-metric} is independent of time, there exists a time-like Killing vector: $\zeta^{\mu}=\left(1,0,0,0\right)^{T}$. This implies $\nabla_{(\mu}\zeta_{\nu)}=0$. We denote $v^{\mu}=\frac{\mathrm{d}x^{\mu}}{\mathrm{d}\tau}$ as the four-velocity of an  observer following a time-like geodesic, with $\tau$ being its proper time. Therefore, $v^{\mu}\nabla_{\mu}v^{\nu}=0$ on the geodesic. Consider the scalar quantity, $\zeta_{\mu}v^{\mu}= -f\left(r\right)\frac{\mathrm{d}t}{\mathrm{d}\tau}$. It can be checked that $\frac{\mathrm{d}}{\mathrm{d}\mathrm{\tau}}\left(\zeta_{\mu}v^{\mu}\right)=0$, implying that $f\left(r\right)\frac{\mathrm{d}t}{\mathrm{d}\tau}:= E$ is a conserved quantity. 

A radial time-like geodesic can then be given by 
\begin{equation}
	\left(\frac{\mathrm{d}r}{\mathrm{d}\tau}\right)^{2} = E^{2}-f\left(r\right)~.
\end{equation}
If the right hand side vanishes at any value of $r$, then the geodesic has a turning point. If the turning point is near the black hole singularity, $r\approx 0$ then $f\left(r\right)\sim -\frac{1}{r}$ and $E^{2}\rightarrow -\infty$. In this limit,
\begin{equation}
	\left\vert\frac{\mathrm{d}r}{\mathrm{d}t}\right\vert = \frac{f\left(r\right)\sqrt{E^{2}-f\left(r\right)}}{E} \sim f\left(r\right)~,
\end{equation}
the time-like geodesic tends to a null-geodesic. This is referred to as ``bouncing geodesic'' in \cite{Grozdanov:2026cut}. Therefore, classically a turning point close to the singularity exists for imaginary energy 
\begin{equation}
	E=i \widetilde{E}~, \qquad \widetilde{E}\in \mathbb{R}~.
\end{equation}
At large energy as the time-like geodesic approaches the null limit, we can define bouncing time as 
\begin{equation}
	t_{\ast} = -2\int_{r}^{0}\frac{\mathrm{d}r}{f\left(r\right)} = 2\left[z\left(r=0\right)-z\left(r\right)\right]~.
\end{equation} 
We will use the contour prescription given in Eq.\eqref{contour-prescription} for the above integral. Bouncing time can be interpreted as the time taken by a light ray to reach the black hole singularity from a time-like surface at $r$ and return to the same point. 

\section{Scalar field probe}
\label{Sec:scalar}
We consider a probe scalar field with mass, $m$ minimally coupled to gravity in the Schwarzschild dS background, 
\begin{equation}
	S= -\frac{1}{2}\int\mathrm{d}^{4}x\; \sqrt{-g}\biggl\{g^{\mu\nu}\partial_{\mu}\phi\partial_{\nu}\phi+m^{2}\phi^{2}\biggr\}~.
\end{equation}
 Equation of motion for the scalar field is
\begin{equation}
	\begin{aligned}
		 \Biggl\{\frac{1}{\sqrt{-g}}\partial_{\mu}\left(\sqrt{-g}g^{\mu\nu}\partial_{\nu}\right)-m^{2}\Biggr\}\phi &=0\\
		\Rightarrow \quad \Biggl\{-\frac{1}{f\left(r\right)}\partial_{t}^{2}+\frac{1}{r^{2}}\partial_{r}\left(r^{2}f\left(r\right)\partial_{r}\right)+\frac{1}{r^{2}}\nabla_{\mathbb{S}^{2}}^{2}-m^{2}\Biggr\}\phi &= 0~.
	\end{aligned}
\end{equation}
We use the following separation of variables, 
\begin{equation}
	\phi\left(x\right)=\frac{1}{2\pi r}\sum_{\ell=0}^{\infty}\sum_{\mu=-\ell}^{\ell}\int_{-\infty}^{\infty}\mathrm{d}\omega\; e^{-i\omega t}Y_{\ell \mu}\left(\theta, \phi\right)\psi_{\omega\ell \mu}\left(r\right)~,
\end{equation}
with $\nabla_{\mathbb{S}^{2}}^{2}Y_{\ell \mu}=-\ell\left(\ell+1\right)Y_{\ell \mu}$, $\forall \mu\in\{-\ell, \cdots,\ell\}$, to obtain 
\begin{equation}
\label{psi_eq}
	\Biggl\{f\left(r\right)\partial_{r}^{2}+ f'\left(r\right)\partial_{r}+\frac{\omega^{2}}{f\left(r\right)}-\frac{f'\left(r\right)}{r}-\frac{\ell\left(\ell+1\right)}{r^{2}}-m^{2}\Biggr\}\psi_{\omega\ell \mu}\left(r\right)=0~.
\end{equation}
First derivative term can be removed by going to tortoise coordinate. Radial equation takes the form
\begin{equation}\label{scalar-eom}
	\left[\partial_{z}^{2}+\omega^{2}-V\left(r\left(z\right)\right)\right]\psi\left(r\left(z\right)\right)=0~,
\end{equation}
where the potential is given by\footnote{Note that we have omitted the subscripts in $\psi$ and we will adhere to this notation. }
\begin{equation}
	V\left(r\right)=f\left(r\right)\left[m^{2}-\frac{2\Lambda}{3}+\frac{\ell\left(\ell+1\right)}{r^{2}}+\frac{2GM}{r^{3}}\right]~.
\end{equation}
Potential vanishes at the two horizons. Therefore solution becomes plane waves, $\psi\left(z\right)\sim e^{\pm i\omega z}$ at the horizons. 

\subsection{Boundary conditions} 
We are interested in analyzing quasi-normal modes of the retarded scalar Green's function in Schwarzschild dS space with a cavity wall placed between the black hole and cosmological horizons \cite{Grozdanov:2026ktq}. We choose the cavity wall to be placed at the static sphere and impose the following boundary conditions on the wave function:
\begin{itemize}
	\item  Ingoing at the black hole horizon. This implies $\psi\left(z\rightarrow \infty\right)\sim e^{i\omega z}$.
	\item Dirichlet condition at the cavity wall, \textit{i.e.} $\psi\left(z=0\right)=0$.
\end{itemize} 
We have to then allow for both ingoing and outgoing plane waves to be present at the cosmological horizon, and we can write the solution as
\begin{equation}
	\psi\left(z\rightarrow-\infty\right) \sim D_{1}e^{i\omega z} + D_{2}e^{-i\omega z}~.
\end{equation}
$D_{1}$ and $D_{2}$ are some constants and they are related due to the boundary conditions. 

\subsection{Asymptotic QNM}
\begin{figure}[h!]
 \centering
	\includegraphics[scale=0.7]{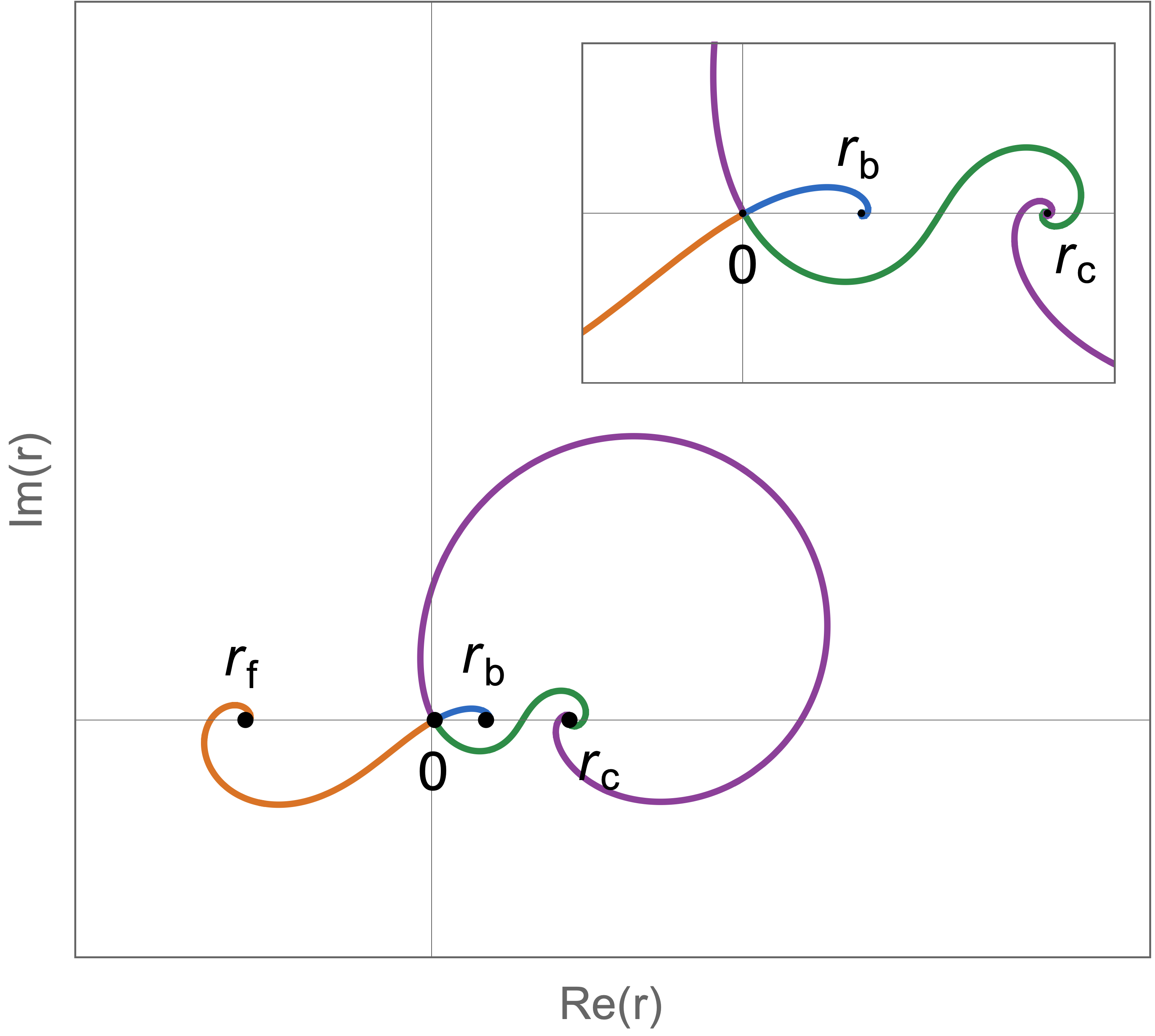}
    \caption{Representative graph showing Stokes curves for asymptotic quasi-normal modes in the complex radial coordinate plane. Black hole horizon and cosmological horizon are denoted by  $r_{b}$ and $r_{c}$ respectively. Negative root of $f\left(r\right)=0$ is marked $r_{f}$. A zoomed region including the static patch has been shown in the inset.}
      \label{Fig:StokesGeom}
\end{figure}

Numerical analysis of quasi-normal modes at large overtone values shows that the Stokes geometry\footnote{We use the terminology used in \cite{Andersson:2003fh,Cardoso:2004up}. Stokes curve refers to the branch having $\operatorname{Im}\left(\omega\xi\right)=0$.} has the form as depicted in Fig.(\ref{Fig:StokesGeom}). Here, one Stokes curve connects  $r=0$ and the black hole horizon. This is drawn in blue. Another one, drawn in green, passes between the singularity and the cosmological horizon. More details on the Stokes geometry are presented in Appendix (\ref{App:Stokes}).
We will map the solutions of Eq.\eqref{scalar-eom}, valid near the  horizons, to each other by tracing along the above mentioned Stokes curves and impose the Dirichlet boundary condition at the cavity wall to derive quantization condition of the asymptotic quasi-normal modes.

\subsection{Solution near singularity}
Around $r=0$, $f\sim-\frac{2GM}{r}$ and potential has the leading form $V\sim -\frac{4G^{2}M^{2}}{r^{4}}$.  Let us denote $z\left(r=0\right)$ by $z_{0}$. Then $2z_{0}$ is the bouncing time corresponding to the cavity wall. In the neighborhood of $z_{0}$, we find
\begin{equation}
	\begin{aligned}
			\int_{z_{0}}^{z}\mathrm{d}z & = -\int_{0}^{r}\frac{\mathrm{d}r}{f\left(r\right)} \\
			& = \int_{0}^{r}\mathrm{d}r \frac{r}{2GM}\left(1-\frac{r}{2GM}+\frac{\Lambda r^3}{6GM}\right)^{-1}~. 
	\end{aligned}
\end{equation}
We can expand the integrand in small $r$ and integrate term by term. This leads to 
\begin{equation}\label{xi-expansion}
	\xi := z-z_{0}=\frac{r^2}{4GM}+\frac{r^3}{12\left(GM\right)^2}+\frac{r^4}{32\left(GM\right)^3}+\mathcal{O}\left(r^{5}\right)~.
\end{equation}
Stokes curves trace out the trajectories in complex $r$-plane where the solutions remain purely oscillatory, \textit{i.e.} $\operatorname{Im}\left(\omega\xi\right)=0$. For the asymptotic quasi-normal modes, let us consider $\omega=|\omega|\exp\left(i\phi\right)$ with $-\pi<\phi<0$. At leading order $\xi\sim r^{2}$. If $\theta$ denotes the argument of the Stokes curves, we obtain
\begin{equation}\label{stokes-angle}
	\theta = -\frac{\phi}{2}+\frac{k\pi}{2}~, \qquad k\in\{0,1,2,3\}~.
\end{equation}
$k$ values correspond to blue, violet, orange and green curves in Fig.(\ref{Fig:StokesGeom}) respectively. Therefore on the blue curve $\omega\xi$ is positive, whereas on the green curve it is negative.

Inverting the relation in Eq.(\ref{xi-expansion}), we can write
	\begin{equation}\label{r-xi-expansion}
		r=2\sqrt{GM\xi}-\frac23\xi+\frac{\xi^{\frac{3}{2}}}{18\sqrt {GM}}+\mathcal{O}\left(\xi^{2}\right).
	\end{equation}
Scalar field equation of motion in Eq.\eqref{scalar-eom} takes the form
\begin{equation}\label{xi-eom}
	\left[\frac{\mathrm{d}^{2}}{\mathrm{d}\xi^{2}}+\omega^{2}-V\left(\xi\right)\right]\psi =0~,
\end{equation}
with the potential given by
\begin{equation}\label{potentialexpansion}
		V\left(\xi\right)=-\frac1{4\xi^2}
		-\frac{3\ell\left(\ell+1\right)+1}{12\sqrt {GM}}\xi^{-\frac{3}{2}} + \mathcal{O}\left(\xi^{-\frac{1}{2}}\right)~.
\end{equation}

\subsection{Leading order solution}
Let us denote the leading solution to be $\psi^{(0)}$. Keeping only the $\xi^{-2}$ term in the potential expansion given in Eq.\eqref{potentialexpansion}, we get 
\begin{equation}\label{leading-scalar-eq}
	\left[\frac{\mathrm{d}^{2}}{\mathrm{d}\xi^{2}}+\omega^{2} + \frac{1}{4\xi^{2}}\right]\psi^{(0)} =0~.
\end{equation}
Performing a change of variable $u=\omega\xi$, followed by the substitution $\psi^{(0)}=\sqrt{u}\varphi\left(u\right)$, we obtain the following Bessel differential equation
\begin{equation}
	\varphi''\left(u\right)+\frac{1}{u}\varphi'\left(u\right)+\varphi\left(u\right)=0~.
\end{equation}
This leads to the general solution,
\begin{equation}\label{gen-sol}
	\psi^{(0)} = \sqrt{2\pi\omega\xi}\left[A J_{0}\left(\omega\xi\right)+BY_{0}\left(\omega\xi\right)\right]~,
\end{equation}
where $J_{\nu}\left(x\right)$ and $Y_{\nu}\left(x\right)$ are the Bessel function of first kind and second kind respectively. Therefore, near $r=0$ the leading order solution behaves as
\begin{equation}
	\psi^{(0)}\left(r\rightarrow 0\right)\sim\sqrt{\frac{\pi\omega}{2GM}}r\left[A+\frac{2B}{\pi}\Biggl\{\log\left(\frac{\omega r^{2}}{8GM}\right)+\gamma_{E}\Biggr\}\right] + \cdots~.
\end{equation}

\paragraph{Connecting to black hole horizon:}
Near the black hole horizon the solution is proportional to $e^{i\omega z}$ and can be expressed as 
\begin{equation}\label{horizon-sol}
	\psi = \mathcal{N}_{b}e^{i\omega\xi}~.
\end{equation} 
$\mathcal{N}_{b}$ is the normalization constant. On the blue Stokes curve in Fig.(\ref{Fig:StokesGeom}), connecting $r=0$ to $r=r_{b}$,  $u=\omega\xi>0$ . From the asymptotic form of the Bessel functions,
\begin{equation}
	\begin{aligned}
		J_{0}\left(u\right)&=\sqrt{\frac{2}{\pi u}}
		\left[\cos\left(u-\frac{\pi}{4}\right)+\mathcal{O}\left(u^{-1}\right)\right]~,\\
		Y_{0}\left(u\right)&=\sqrt{\frac{2}{\pi u}}
		\left[\sin\left(u-\frac{\pi}{4}\right)+\mathcal{O}\left(u^{-1}\right)\right]~,
	\end{aligned}
	 \qquad u\gg 1
\end{equation}
and using the general solution given in Eq.\eqref{gen-sol}, we obtain  
\begin{equation}
	\psi^{(0)}\sim \left(A-i B\right)e^{-i\frac{\pi}{4}} e^{i\omega\xi} + \left(A+i B\right)e^{i\frac{\pi}{4}}e^{-i\omega\xi}~, \qquad \omega\xi\gg 1~.
\end{equation}
Comparing this with Eq.\eqref{horizon-sol} yields the following condition,
\begin{equation}
	A \propto\frac{1}{2}\mathcal{N}_{b}e^{i\frac{\pi}{4}}~, \qquad B = i A~.
\end{equation}
Hence, the general solution on this specific Stokes curve reduces to 
\begin{equation}\label{positiveusol}
	\psi^{(0)}=\mathcal{N}_{b}\sqrt{\frac{\pi\omega\xi}{2}}e^{i\frac{\pi}{4}}H^{(1)}_{0}\left(\omega\xi\right)~,
\end{equation}
up to some overall normalization constant. Here $H^{(1)}_{\nu}\left(z\right)=J_{\nu}\left(z\right)+i Y_{\nu}\left(z\right)$ is the Hankel function of the first kind. 

\paragraph{Connecting to cosmological horizon:}
The green coloured Stokes curve in Fig.(\ref{Fig:StokesGeom}), passing through $r=0$ and the cosmological horizon, is rotated by an angle $\frac{\pi}{2}$ in the clockwise direction with respect to the blue Stokes curve\footnote{Another choice of rotation could have been $\frac{3\pi}{2}$ in anticlockwise direction. This would lead to change in $\operatorname{Arg}\left(\omega\xi\right)$ by $3\pi$. This leads to landing on a different Riemann sheet and we should account for the corresponding monodromy contribution in this case.}. Thus $\omega\xi$ changes sign and becomes negative on the green curve,
\begin{equation}
	\omega\xi = |\omega\xi|e^{-i\pi}~.
\end{equation}
We have to account for the change of this phase while going from blue to green Stokes curve \cite{Motl:2003cd}. Continuing the solution in Eq.(\ref{positiveusol}) to the green branch gives 
\begin{equation}\label{negativeusol}
	\begin{aligned}
		\psi^{(0)} &=\mathcal{N}_{b}\sqrt{\frac{\pi|\omega\xi|}{2}}e^{-i\frac{\pi}{4}}\left[J_{0}\left(|\omega\xi|e^{-i\pi}\right)+i Y_{0}\left(|\omega\xi|e^{-i\pi}\right)\right]\\
		& = \mathcal{N}_{b}\sqrt{\frac{\pi|\omega\xi|}{2}}e^{-i\frac{\pi}{4}}\left[J_{0}\left(|\omega\xi|\right) + i \biggl\{Y_{0}\left(|\omega\xi|\right)-2i J_{0}\left(|\omega\xi|\right)\biggr\}\right]\\
		& = \mathcal{N}_{b}\sqrt{\frac{\pi|\omega\xi|}{2}}e^{-i\frac{\pi}{4}}\left[3 J_{0}\left(|\omega\xi|\right)+i Y_{0}\left(|\omega\xi|\right)\right]~.
	\end{aligned}
\end{equation}
Now we take the asymptotic limit of Eq.\eqref{negativeusol}, 
\begin{equation}
	\psi^{(0)}\sim \mathcal{N}_{b}\left[e^{i\omega\xi}-2i e^{-i\omega\xi}\right]~, \qquad |\omega\xi|\gg 1~.
\end{equation}
Here we have used $|\omega\xi|=-\omega\xi$. This can be matched with the solution near cosmological horizon,
\begin{equation}
	\begin{aligned}
		\psi & = D_{1}e^{i\omega z}+D_{2}e^{-i\omega z}~, \qquad z\rightarrow -\infty \\
		& = D_{1}e^{i\omega z_{0}}e^{i\omega\xi}+D_{2}e^{-i\omega z_{0}}e^{-i\omega \xi}~.
	\end{aligned}
\end{equation}
Then the following relations are obtained,
\begin{equation}\label{cosmo-asyymptotic}
	D_{1} \propto \mathcal{N}_{b}e^{-i\omega z_{0}}~, \qquad D_{2}\propto -2i\mathcal{N}_{b}e^{i\omega z_{0}}~.
\end{equation}
We also have the Dirichlet boundary condition at the cavity wall, which gives
\begin{equation}
	\psi\left(z=0\right)=0 \qquad \Rightarrow \qquad e^{-i\omega z_{0}}-2i e^{i\omega z_{0}}=0~.
\end{equation}
The last equality gives $e^{2i\omega z_{0}}=-\frac{i}{2}$ from which we can derive a quantization condition for the asymptotic quasinormal modes,
\begin{equation}\label{leading-asymp-QNM}
	\omega z_{0} = \pi n - \frac{\pi}{4} + \frac{i}{2}\ln 2~, \qquad n\in\mathbb{N}~, \quad n\gg 1~.
\end{equation}

\section{Beyond sub-leading order expansion}
\label{Sec:subleading}
In this section we find corrections to the asymptotic expansion of the large quasi-normal modes to Eq.\eqref{leading-asymp-QNM}. There are two sources from which correction terms are contributed - higher order terms in the expansion of the potential near black hole singularity and sub-leading terms appearing in the phase of the WKB wave-function. 

\subsection{Potential correction}
Here the sub-leading term at $\mathcal{O}\left(\xi^{-\frac{3}{2}}\right)$ in Eq.(\ref{potentialexpansion}) is included in the analysis. Taking $u=\omega\xi$, Eq.\eqref{xi-eom} to sub-leading order becomes
\begin{equation}\label{subleading-eom}
	\left[\frac{\mathrm{d}^{2}}{\mathrm{d}u^{2}}+1+\frac{1}{4u^{2}}+\frac{\alpha}{\sqrt{\omega}u^{\frac{3}{2}}}\right]\psi\left(u\right)=0~, \qquad \alpha = \frac{3\ell\left(\ell+1\right)+1}{12\sqrt {GM}}~.
\end{equation}
Let us consider $\mathcal{L}_{0}:=\frac{\mathrm{d}^{2}}{\mathrm{d}u^{2}}+1+\frac{1}{4u^{2}}$. The homogeneous differential equation, $\mathcal{L}_{0}\psi=0$ has the solutions 
\begin{equation}\label{homogeneousbasis}
	p\left(u\right)= \sqrt{\frac{\pi u}{2}}J_{0}\left(u\right)~, \qquad q\left(u\right)=\sqrt{\frac{\pi u}{2}}Y_{0}\left(u\right)~.
\end{equation}
For large $\omega$, we can solve for $\mathcal{L}_{0}\psi=-\frac{\alpha}{\sqrt{\omega}}u^{-\frac{3}{2}}\psi$ perturbatively. We consider an ansatz for the inhomogeneous solution of the form $\psi\left(u\right) = c_{1}\left(u\right)p\left(u\right)+c_{2}\left(u\right)q\left(u\right)$. Then the problem reduces to finding an evolution equation for the coefficients with respect to $u$. Details of the analysis are presented in Appendix (\ref{App:Volterra}). Plugging the ansatz in Eq.\eqref{subleading-eom}, we obtain a first order differential equation for the coefficients,
\begin{equation}
	\frac{\mathrm{d}}{\mathrm{d}u}\mathbf{c}\left(u\right)=\frac{\alpha}{\sqrt{\omega}}\mathbf{R}\left(u\right)\mathbf{c}\left(u\right)~.
\end{equation}
$\mathbf{R}\left(u\right)$ matrix is defined in Eq.\eqref{evolutionEq}. Writing $\mathbf{c}\left(u\right)=\mathcal{U}\left(u;\frac{\alpha}{\sqrt{\omega}}\right)\mathbf{c}\left(0\right)$, we will get an ordered integral representation of the form $\mathcal{U}\left(u;\frac{\alpha}{\sqrt{\omega}}\right)=\mathcal{P}\exp\left[\frac{\alpha}{\sqrt{\omega}}\int_{0}^{u}\mathbf{R}\left(t\right)\mathrm{d}t\right]$. A closed form expression for $\mathcal{U}$ can be obtained for large $u$,
\begin{equation}\label{connection-matrix-2ndorder}
	\lim_{u\rightarrow\infty}\mathcal{U}\left(u;\frac{\alpha}{\sqrt{\omega}}\right)=\begin{pmatrix}
		1 & 0\\ 0 & 1
	\end{pmatrix} + \frac{\alpha\gamma}{\sqrt{\omega}}\begin{pmatrix}
		-1 & 3\\ -1 & 1
	\end{pmatrix}
	+ \frac{\alpha^{2}\gamma^{2}}{\omega}
	\begin{pmatrix}
		0 & -\frac{2}{3}\\ \frac{2}{3} & -2
	\end{pmatrix}
	+ \mathcal{O}\left(\left(\frac{\alpha}{\sqrt{\omega}}\right)^3\right)~.
\end{equation}
Here $\gamma=\frac{\sqrt{2} \pi ^{\frac{3}{2}} \Gamma \left(\frac{5}{4}\right)}{\Gamma
	\left(\frac{3}{4}\right)^3} $. Coefficients near the singularity can be obtained from Eq.\eqref{coeff-transfer} by inverting the connection matrix,
\begin{equation}
	\begin{aligned}
		\mathbf{c}\left(0\right) &= \mathcal{U}\left(\infty;\frac{\alpha}{\sqrt{\omega}}\right)^{-1}\mathbf{c}\left(\infty\right)\\
		& = \left[\mathbb{I}-\frac{\alpha}{\sqrt{\omega}}H_{1}+\frac{\alpha^{2}}{\omega}\left(H_{1}^{2}-H_{2}\right)+ \mathcal{O}\left(\left(\frac{\alpha}{\sqrt{\omega}}\right)^{3}\right)\right]\mathbf{c}\left(\infty\right)~.
	\end{aligned}
\end{equation}
To go from the blue to green curve, argument of the solution changes by $e^{-i\pi}$. Therefore the basis elements transform as 
\begin{equation}
	\begin{pmatrix}
		p\left(u\right) & q\left(u\right)
	\end{pmatrix} \rightarrow \begin{pmatrix}
	p\left(u\right) & q\left(u\right)
	\end{pmatrix} \begin{pmatrix}
	-i & -2\\ 0 & -i
	\end{pmatrix}~.
\end{equation}
 This implies that the coefficients on the green Stokes curve are related to those on the blue curve by 
\begin{equation}
	\mathbf{c}'\left(0\right)=T\mathbf{c}\left(0\right)~, \qquad T= \begin{pmatrix}
		-i & -2\\ 0 & -i
	\end{pmatrix}~.
\end{equation}
Since $\omega\xi$ changes sign between the two curves, let us take $s=-\omega\xi>0$ on the green branch. Then Eq.\eqref{subleading-eom} becomes
\begin{equation}
	\left[\frac{\mathrm{d}^{2}}{\mathrm{d}s^{2}}+1+\frac{1}{4s^{2}}-\frac{i\alpha}{\sqrt{\omega}s^{\frac{3}{2}}}\right]\psi\left(s\right)=0
\end{equation}
Therefore, $\alpha\rightarrow -i\alpha$ on this Stokes curve and the connection formula becomes $\mathbf{c}'\left(u\right)=\mathcal{U}\left(u;-\frac{i\alpha}{\sqrt{\omega}}\right)\mathbf{c}\left(0\right)$. Finally, coefficients of the asymptotic solutions are given by
\begin{equation}
	\mathbf{c}'\left(\infty\right)= \mathcal{U}\left(\infty;-\frac{i\alpha}{\sqrt{\omega}}\right)\;T\; \mathcal{U}\left(\infty;\frac{\alpha}{\sqrt{\omega}}\right)^{-1}\mathbf{c}\left(\infty\right)~.
\end{equation}
Matching the solutions to their respective forms near the horizons and imposing the wall boundary condition at $z=0$, we can derive the following quantization condition, 
\begin{equation}\label{potentialcorr}
	\omega z_{0} = \pi n-\frac{\pi}{4}+\frac{i}{2}\log 2+\frac{1+i}{2}\frac{\alpha\gamma}{\sqrt{\omega}}-\frac{1}{6}\frac{\alpha^{2}\gamma^{2}}{\omega} + \mathcal{O}\left(\frac{\alpha^{3}}{\omega^{\frac{3}{2}}}\right)~, \qquad n\in\mathbb{N}~, \quad n\gg 1~.
\end{equation}
Coefficient of $\omega^{-\frac{1}{2}}$ term is analogous to the $n^{-\frac{1}{2}}$ correction in case of Schwarzschild black hole quasi-normal modes found in \cite{Musiri:2003bv,Kao:2007vr}.

\subsection{Wavefunction correction}
For $\omega^{2}\gg V\left(z\right)$, leading order solution to Eq.\eqref{scalar-eom} behaves as $e^{\pm i\omega z}$. To find the sub-leading corrections, let us assume WKB ansatz $\psi = e^{i\omega  S\left(\omega,z\right)}$, with 
\begin{equation}
	S\left(\omega,z\right)= S^{(0)}\left(z\right) + \frac{S^{(1)}\left(z\right)}{\omega} + \frac{S^{(2)}\left(z\right)}{\omega^{2}} + \cdots~, \qquad |\omega|\gg 1~.
\end{equation}
Plugging this ansatz in Eq.\eqref{scalar-eom} leads to 
\begin{equation}
	S^{(0)'}\left(z\right) = \pm 1~, \qquad S^{(1)'}\left(z\right)=0~, \qquad S^{(2)'}\left(z\right)=\mp \frac{1}{2}V\left(z\right)~.
\end{equation}
Hence, phase of the wave function at far region receives correction of the following form 
\begin{equation}
	\psi \propto \exp\left[\pm i \omega\left(z-z_{0}\right)\mp \frac{i}{2\omega}\int_{z_{0}}^{z}V\left(z\right)\mathrm{d}z\right]~.
\end{equation}
Using $f\left(r\right)\mathrm{d}z=-\mathrm{d}r$, we can write
\begin{equation}
	\int_{z_{0}}^{0}V\left(z\right)\mathrm{d}z = -\lim_{r\rightarrow 0}\int_{r}^{r_{\mathcal{O}}}\left[m^{2}-\frac{2\Lambda}{3}+\frac{\ell\left(\ell+1\right)}{r^{2}}+\frac{2GM}{r^{3}}\right] \mathrm{d}r~.
\end{equation}
Note that $r=0$ produces a divergence in the phase factor and therefore the lower limit of the integration has to be regulated. We can substitute $\xi$ in terms of $r$ near the singularity by using the expansion given in Eq.\eqref{r-xi-expansion}. Above equation then takes the form 
\begin{equation}\label{phase-potential}
	\begin{aligned}
		\int_{z_{0}}^{0}V\left(z\right)\mathrm{d}z 
		=&  \lim_{\xi\rightarrow 0}\left[\ell\left(\ell+1\right)\biggl\{\left(\frac{\Lambda}{3GM}\right)^{\frac{1}{3}}-\frac{1}{6GM}\biggr\}-\left(m^{2}-\Lambda\right)\left(\frac{3GM}{\Lambda}\right)^{\frac{1}{3}}\right.\\
		& \qquad \left. -\frac{5}{72GM}-\frac{3\ell\left(\ell+1\right)+1}{6\sqrt{GM\xi}}-\frac{1}{4\xi}\right]~.
	\end{aligned}
\end{equation}
Last two terms diverge as $\xi\rightarrow 0$. Very close to the singularity, WKB approximation is subtle because the potential itself is large. This can also be understood from Eq.\eqref{subleading-eom} - plane wave approximation is valid away from $\xi=0$. Therefore, the lower limit of the above integration should be moved to a non-zero value, $\xi_{0}$. As this additional phase factor, $\exp\left[\pm\frac{i}{2\omega}\biggl\{\frac{3\ell\left(\ell+1\right)+1}{6\sqrt{GM\xi_{0}}}+\frac{1}{4\xi_{0}}\biggr\}\right]$ is independent of the location of the wall, it can be absorbed in the respective coefficient term.

Regular terms in the phase factor will affect the wave function at the wall placed at $z=0$. Vanishing of the wave function at $z=0$ then implies
\begin{equation}
	D_{1}e^{-\frac{i}{2\omega}\left[\int_{z_{0}}^{0}V\left(z\right)\mathrm{d}z\right]_{\text{regular}}} + D_{2}e^{\frac{i}{2\omega}\left[\int_{z_{0}}^{0}V\left(z\right)\mathrm{d}z\right]_{\text{regular}}} = 0~.
\end{equation}
Here the regular expressions are defined by removing $\frac{3\ell\left(\ell+1\right)+1}{6\sqrt{GM\xi}}+\frac{1}{4\xi}$ terms from Eq.\eqref{phase-potential}. Invoking Eq.\eqref{cosmo-asyymptotic} we can obtain
\begin{equation}\label{lambda-contribution}
	\begin{aligned}
		\omega z_{0}=& \pi n-\frac{\pi}{4}+\frac{i}{2}\ln 2-\frac{1}{2\omega}\left[\ell\left(\ell+1\right)\biggl\{\left(\frac{\Lambda}{3GM}\right)^{\frac{1}{3}}-\frac{1}{6GM}\biggr\}\right.\\
		& \qquad   \left.-\left(m^{2}-\Lambda\right)\left(\frac{3GM}{\Lambda}\right)^{\frac{1}{3}}-\frac{5}{72GM}\right] + \mathcal{O}\left(\omega^{-\frac{3}{2}}\right)~, \qquad n\in\mathbb{N}~, \quad n\gg 1~. 
	\end{aligned}
\end{equation}
So the cosmological constant dependent sub-leading phase factor corrects the asymptotic quasi-normal modes at  $\frac{1}{n}$ order for large $n$. 

\paragraph{Final expression:} Combining the sub-leading corrections from Eq.\eqref{potentialcorr} and Eq.\eqref{lambda-contribution}, we obtain the following asymptotic form of the large quasi-normal modes to $\omega^{-1}$ order,
\begin{equation}
	\begin{aligned}
		\omega z_{0} =& \pi n-\frac{\pi}{4}+\frac{i}{2}\log 2+\frac{\left(1+i\right)\alpha\gamma}{2\sqrt{\omega}} - \frac{1}{2\omega}\left[\frac{\alpha^{2}\gamma^{2}}{3}+\ell\left(\ell+1\right)\biggl\{\left(\frac{\Lambda}{3GM}\right)^{\frac{1}{3}}-\frac{1}{6GM}\biggr\}\right.\\
		& \qquad  \left.-\left(m^{2}-\Lambda\right)\left(\frac{3GM}{\Lambda}\right)^{\frac{1}{3}}-\frac{5}{72GM} \right] + \mathcal{O}\left(\omega^{-\frac{3}{2}}\right)~, \quad n\in\mathbb{N}~, \quad |\omega|\gg 1~,
	\end{aligned}
\end{equation}
with $\alpha = \frac{3\ell\left(\ell+1\right)+1}{12\sqrt {GM}}$ and $\gamma=\frac{\sqrt{2} \pi ^{\frac{3}{2}} \Gamma \left(\frac{5}{4}\right)}{\Gamma
	\left(\frac{3}{4}\right)^3} $. Since $n$ is large, we can write
	\begin{equation}\label{QNM-exp}
		\omega_{n} z_{0}= \pi n+A_{0} + \frac{A_{-\frac{1}{2}}}{\sqrt{n}}+\frac{A_{-1}}{n} + \mathcal{O}\left(n^{-\frac{3}{2}}\right)~, \qquad n\gg 1
	\end{equation}
	 with the coefficients given by 
	\begin{equation}
		\begin{aligned}
			A_{0} & = -\frac{\pi}{4}+\frac{i}{2}\log 2~,\\
			A_{-\frac{1}{2}}& = \frac{\left(1+i\right)\alpha\gamma}{2}\sqrt{\frac{z_{0}}{\pi}}~,\\
			A_{-1} & = - \frac{z_{0}}{2\pi}\left[\frac{\alpha^{2}\gamma^{2}}{3}+\ell\left(\ell+1\right)\biggl\{\left(\frac{\Lambda}{3GM}\right)^{\frac{1}{3}}-\frac{1}{6GM}\biggr\}\right.\\
			& \qquad  \left.-\left(m^{2}-\Lambda\right)\left(\frac{3GM}{\Lambda}\right)^{\frac{1}{3}}-\frac{5}{72GM} \right]~. 
		\end{aligned}
	\end{equation}
    Throughout the calculation we take the branch with  $\operatorname{Im}\left(z_{0}\right)=\frac{\pi}{f'\left(r_{b}\right)}$.

\section{Numerical analysis}
\label{Sec:Num}
        To independently determine the quasinormal mode frequencies without relying on the WKB or Stokes approximation, we solve the exact radial differential equation, Eq.\eqref{psi_eq} directly using a Frobenius series expansion \cite{Leaver1985}  around the black hole horizon $r = r_b$. In this section we will set $G=1$.
        
        To make the boundary condition manifest at the black hole horizon and remove the essential oscillatory singularity at $r_b$, we factor out the ingoing phase of the form
        \begin{equation}
            \psi(r) = e^{i\omega z(r)} h(r)~.
        \end{equation}
        Substituting this into Eq.~\eqref{psi_eq}, we obtain the regularized differential equation for the envelope function $h(r)$:
        \begin{equation}
        \label{psi_eq_r}
            f(r) h''(r) + \left[ f'(r) - 2i\omega \right] h'(r) - U(r) h(r) = 0~,
        \end{equation}
        where the reduced potential $U(r) \equiv \frac{V\left(r\right)}{f\left(r\right)}$ is given by,
        \begin{equation}
            U(r) = m^2 + \frac{\ell(\ell+1)}{r^2} + \frac{f'(r)}{r} = m^2 + \frac{L}{r^2} + \frac{2M}{r^3} - \frac{2\Lambda}{3}.
        \end{equation}
        Here we have defined $L:=\ell\left(
        \ell+1\right)$.
        In terms of the envelope function $h(r)$, the physical boundary conditions become:
        \begin{equation}
        \label{boundcond2}
            h(r) \sim \begin{cases}
            1 & \text{at } r = r_b \quad (\text{regular ingoing condition at the black hole horizon}), \\
            0 & \text{at } r = r_\mathcal{O}  \quad (\text{reflecting Dirichlet wall at the static sphere}).
        \end{cases}
        \end{equation}
        To map the active physical cavity $\left[r_b, r_\mathcal{O}\right]$ onto a normalized domain, we define a dimensionless coordinate $x \in [0, 1]$ as
        \begin{equation}
        \label{hx_c_k}
            x = \frac{r - r_b}{d}~, \quad d \equiv r_\mathcal{O} - r_b ~, \quad r(x) = r_b + d x~.
        \end{equation}
        The envelope function can then be expanded in a power series,
        \begin{equation}
            h(x) = \sum_{k=0}^{\infty} c_k(\omega) x^k~, \quad \text{with } c_0 = 1~.
        \end{equation}
        Setting $c_0 = 1$ automatically enforces the ingoing plane wave condition at the black hole horizon, $x = 0$ ($r = r_b$).

        In terms of $x$ coordinate the equation of motions becomes,
        \begin{equation}
        \label{hxdiffeq}
        \mathcal{A}(r(x)) h''(x) + d\,\mathcal{B}(r(x)) h'(x) + d^2\,\mathcal{C}(r(x)) h(x) = 0~,
        \end{equation}
        where the polynomial coefficient functions are:
        \begin{align}
        \mathcal{A}(r) &= r^3 f(r) = r^3 - 2Mr^2 - \frac{\Lambda}{3}r^5, \\
        \mathcal{B}(r) &= r^3\left[f'(r) - 2i\omega\right] = 2Mr - \frac{2\Lambda}{3}r^4 - 2i\omega r^3, \\
        \mathcal{C}(r) &= -r^3 U(r) = -2M - Lr - \left( m^2 - \frac{2\Lambda}{3} \right)r^3.
        \end{align}
        Expanding each polynomial in powers of $x = \frac{(r - r_b)}{d}$, we obtain
        \begin{equation}
            \mathcal{A}(r(x)) = \sum_{j=1}^5 \mathcal{A}_j x^j, \quad d\,\mathcal{B}(r(x)) = \sum_{j=0}^4 \mathcal{B}_j x^j, \quad d^2\,\mathcal{C}(r(x)) = \sum_{j=0}^3 \mathcal{C}_j x^j.
        \end{equation}
        Notice that $\mathcal{A}_{0} = \mathcal{A}(r_b) = r_b^3 f\left(r_b\right) = 0$ because $r_b$ is a root of $f\left(r\right)$. Substituting $h(x) = \sum_{k=0}^\infty c_k x^k$ into Eq.~\eqref{hxdiffeq} and collecting the coefficient of $x^k$ gives the explicit recurrence relation:
        \begin{align}
            c_{k+1} = -\frac{1}{(k+1)\left(k \mathcal{A}_1 + \mathcal{B}_0\right)}\Biggl[ &\sum_{j = 2}^5 \mathcal{A}_j\left(k-j+2\right)\left(k-j+1\right)c_{k-j+2} \nonumber \\
            &+ \sum_{j = 1}^4 \mathcal{B}_j\left(k-j+1\right)c_{k-j+1} + \sum_{j= 0}^3 \mathcal{C}_j c_{k-j} \Biggr],
        \end{align}
        where any coefficient $c_m$ with $m < 0$ is set to zero.

        The power series $h(x)$ is centered at the black hole horizon $r = r_b$. The singular points of the differential equation, Eq.\eqref{psi_eq} in the complex $r$-plane are the roots of $f(r) = 0$ (at $r_b, r_c, r_f$) and the curvature singularity at $r = 0$. Since $r_b$ is the expansion center, the nearest singularity is either $r = r_c$ or $r = 0$. Whenever $r = r_c$ is the nearest singularity, $ r = r_{\mathcal{O}}$ always lies inside the disk of convergence. Otherwise, the distance to the nearest other singularity is $|r_b - 0| = r_b$. The distance from the horizon to the cavity wall is $d = r_\mathcal{O} - r_b$. Whenever the ratio $d/r_b$ is strictly less than unity, the Dirichlet wall $r = r_\mathcal{O}$ ($x = 1$) lies entirely within the disk of convergence of the Taylor series. Therefore, this method covers only $0.593 < 3GM\sqrt{\Lambda} < 1$, around $40\%$ of the sub-Nariai range.

        The Dirichlet wall boundary condition requires the field to vanish at $x = 1$. Therefore, 
        \begin{equation}
            F(\omega) \equiv h(x=1) = \sum_{k=0}^{\infty} c_k(\omega) = 0.
        \end{equation}
        In practical numerical computation, the series is truncated at a large cutoff $N$:
        \begin{equation}
            F_N(\omega) \equiv \sum_{k=0}^{N} c_k(\omega) = 0.
        \end{equation}
       The complex eigenfrequencies are computed via the Newton--Raphson method:
        \begin{equation}
            \omega^{(q+1)} = \omega^{(q)} - \frac{F_N(\omega^{(q)})}{F'_N(\omega^{(q)})}.
        \end{equation}
\begin{table}[htbp]
\centering
\footnotesize
\setlength{\tabcolsep}{2.8pt}
\vspace{0.4em}
\begin{tabular}{@{}llcc@{}}
\toprule
\textbf{Parameters} & \textbf{Coeff.} & \textbf{Numerical Fit} & \textbf{Residual} \\ \midrule
$\ell=0, m=0, M=1, \Lambda=0.06$ & $A_{1}$& $3.1415926536 - 3.20\times 10^{-19}\,i$ & $3.20\times 10^{-19}$ \\
& $A_{0}$& $-0.7853981634 + 0.3465735903\,i$ & $1.64\times 10^{-14}$ \\ 
$\ell=2, m=0, M=1, \Lambda=0.06$ & $A_{1}$& $3.1415926536 + 6.29\times 10^{-12}\,i$ & $2.65\times 10^{-11}$ \\
& $A_{0}$& $-0.7853968238 + 0.3465732452\,i$ & $1.38\times 10^{-6}$ \\ 
$\ell=1, m=0, M=1,\Lambda=0.09$& $A_{1}$& $3.1415926536 - 1.37\times 10^{-15}\,i$ & $8.11\times 10^{-15}$ \\
& $A_{0}$& $-0.7853981629 + 0.3465735904\,i$ & $5.05\times 10^{-10}$ \\ 
$\ell=1, m=0, M=1.2, \Lambda=0.06$& $A_{1}$& $3.1415926536 - 3.63\times 10^{-16}\,i$ & $5.78\times 10^{-15}$ \\
& $A_{0}$& $-0.7853981630 + 0.3465735903\,i$ & $3.62\times 10^{-10}$ \\ 
$\ell=1, m=2, M=1, \Lambda=0.06$& $A_{1}$& $3.1415926536 - 4.64\times 10^{-16}\,i$ & $7.12\times 10^{-15}$ \\
& $A_{0}$& $-0.7853981641 + 0.3465735903\,i$ & $7.09\times 10^{-10}$ \\ \bottomrule
\end{tabular}
\caption{Numerical verification of coefficients $A_{1}$ and $A_{0}$.}
\label{tab:A1_A0_verification}
\end{table}

        \subsection{Numerical checks for sub-leading coefficients}

        Having computed high-precision spectra across large overtone numbers $n$ (sampled up to $n = 5000$), the dimensionless phase $\omega_n z_0$ is fitted to the form
        \begin{equation}\label{Num-omega-exp}
            \omega_n z_0 = A_{1} n+A_{0} + \frac{A_{-\frac{1}{2}}}{\sqrt{n}}+\frac{A_{-1}}{n} + \frac{A_{-\frac{3}{2}}}{n^{3/2}} + \dots + \frac{A_{-\frac{7}{2}}}{n^{7/2}}.
        \end{equation}
        The fit is performed over high overtones, in particular $n$ in the range $200-5000$ and retains higher-order buffer terms up to $n^{-7/2}$ to eliminate truncation bias in the lower coefficients. The extracted values for $A_{1},A_{0}, A_{-\frac{1}{2}}, A_{-1}$ are then compared against the analytic formulas derived via Stokes matching in Eq.~\eqref{QNM-exp}. The numerically fitted $A_{1}$ and $A_{0}$ values are given in Table~(\ref{tab:A1_A0_verification}) along with the residuals, which are obtained when subtracted from their respective analytical values from Eq.~(\ref{QNM-exp}).
        
\begin{figure}[h!]
    \centering
    \begin{subfigure}[b]{0.95\textwidth}
        \centering
        \includegraphics[width=\textwidth]{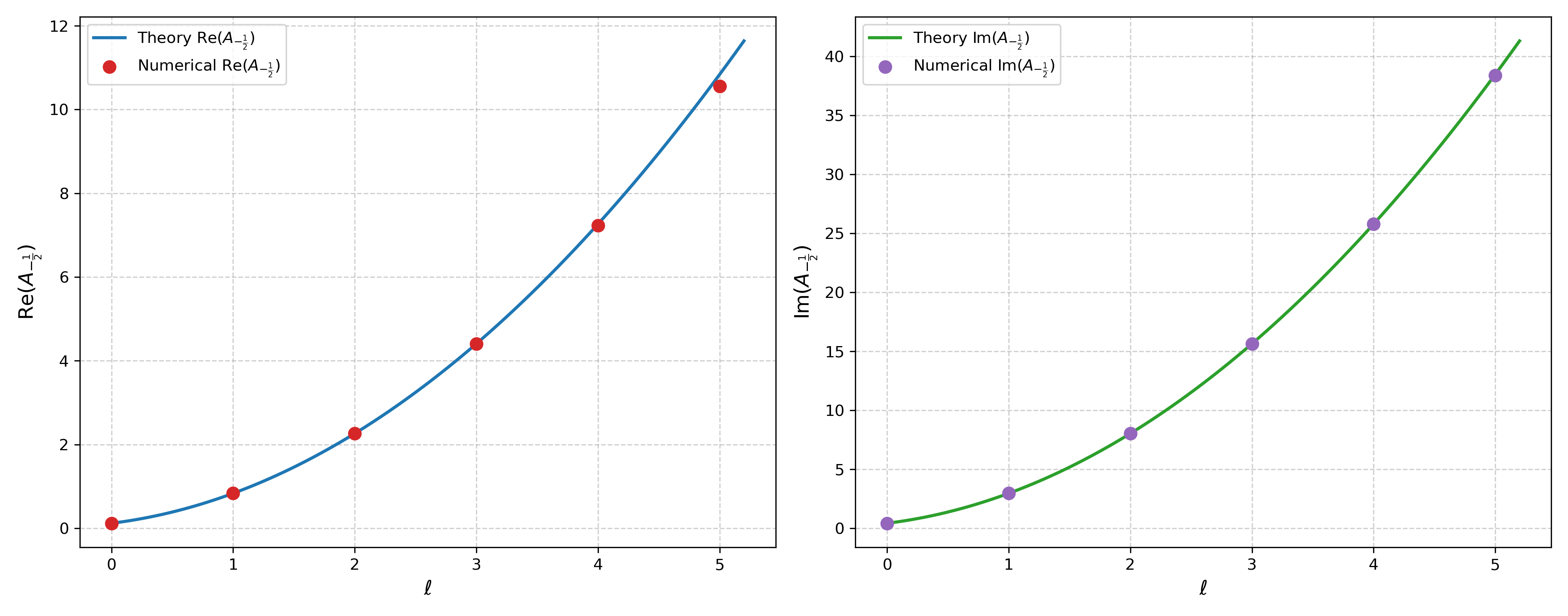}
        \caption{$A_{-\frac{1}{2}}$ vs $\ell$}
        \label{fig:B2_ell}
    \end{subfigure}

    \vspace{0.8em}

    \begin{subfigure}[b]{0.95\textwidth}
        \centering
        \includegraphics[width=\textwidth]{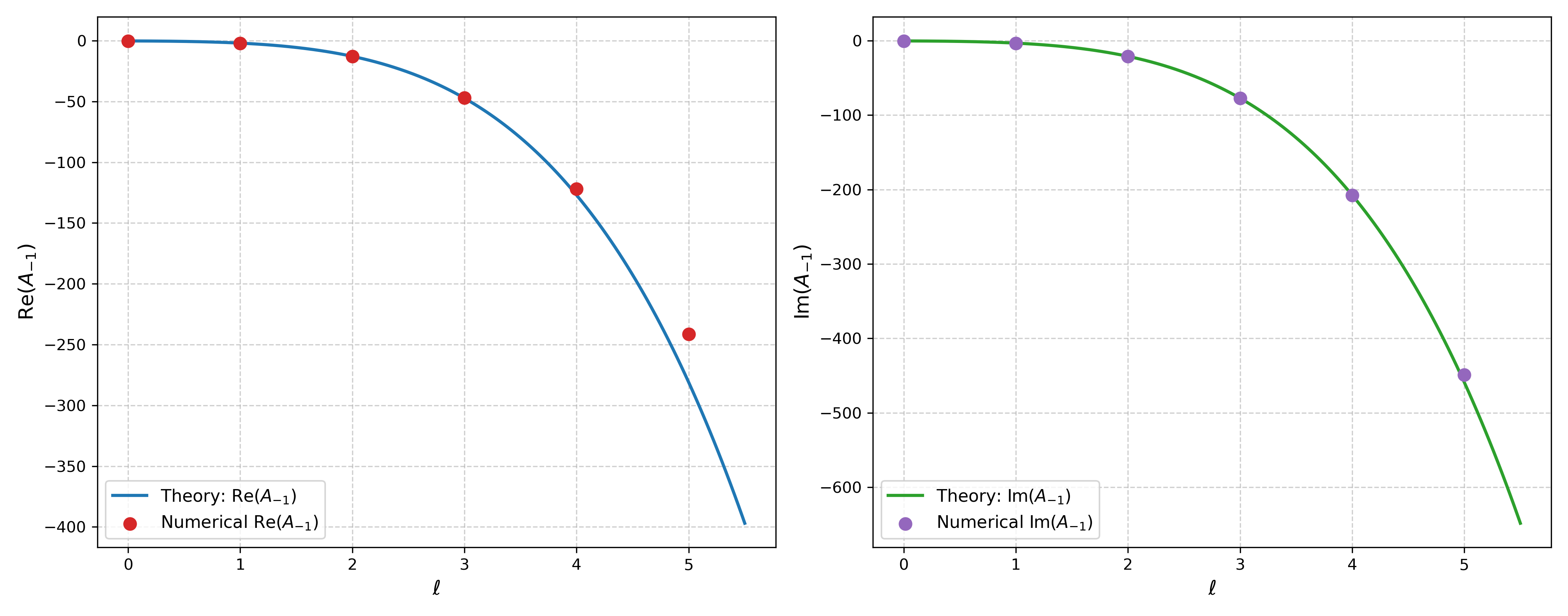}
        \caption{$A_{-1}$ vs $\ell$}
        \label{fig:B3_ell}
    \end{subfigure}
    \caption{Subleading coefficients in the expansion of $\omega_{n}z_{0}$ vs orbital angular momentum $\ell \in \{0, 1, 2, 3, 4, 5\}$ for $M=1, \Lambda=0.06, m=0$. Solid curves represent the theoretical predictions. (a) Displays pure quadratic growth $\propto (3\ell^2+3\ell+1)$. (b) Displays steep quartic growth dominated by $[3\ell(\ell+1)+1]^2$.}
    \label{fig:ell_sweeps}
\end{figure}
In Figure~(\ref{fig:ell_sweeps}), the numerical points (red circles for $\mathrm{Re}$, purple for $\mathrm{Im}$) lie exactly on the continuous analytic curves (blue and green) for smaller values of $\ell$ but for large $\ell = 4$ and $5$ values there are small deviations from the theoretical values. This is because, in Eq.~\eqref{QNM-exp} the terms being studied are $A_{-\frac{1}{2}}/\sqrt{n} \sim (\ell(\ell+1))/\sqrt{n}$ and $A_{-1}/n \sim (\ell(\ell+1))^2/n$. For $\ell = 4$ and $5$ the values $(\ell(\ell+1))/\sqrt{n}$ become order one for the chosen values of $n$ to do the numerical analysis. We have verified that the deviations become increasingly smaller as $n$ is taken over a range of $2000 - 5000$. 
\begin{figure}[h!]
    \centering
    \begin{subfigure}[b]{0.95\textwidth}
        \includegraphics[width=\textwidth]{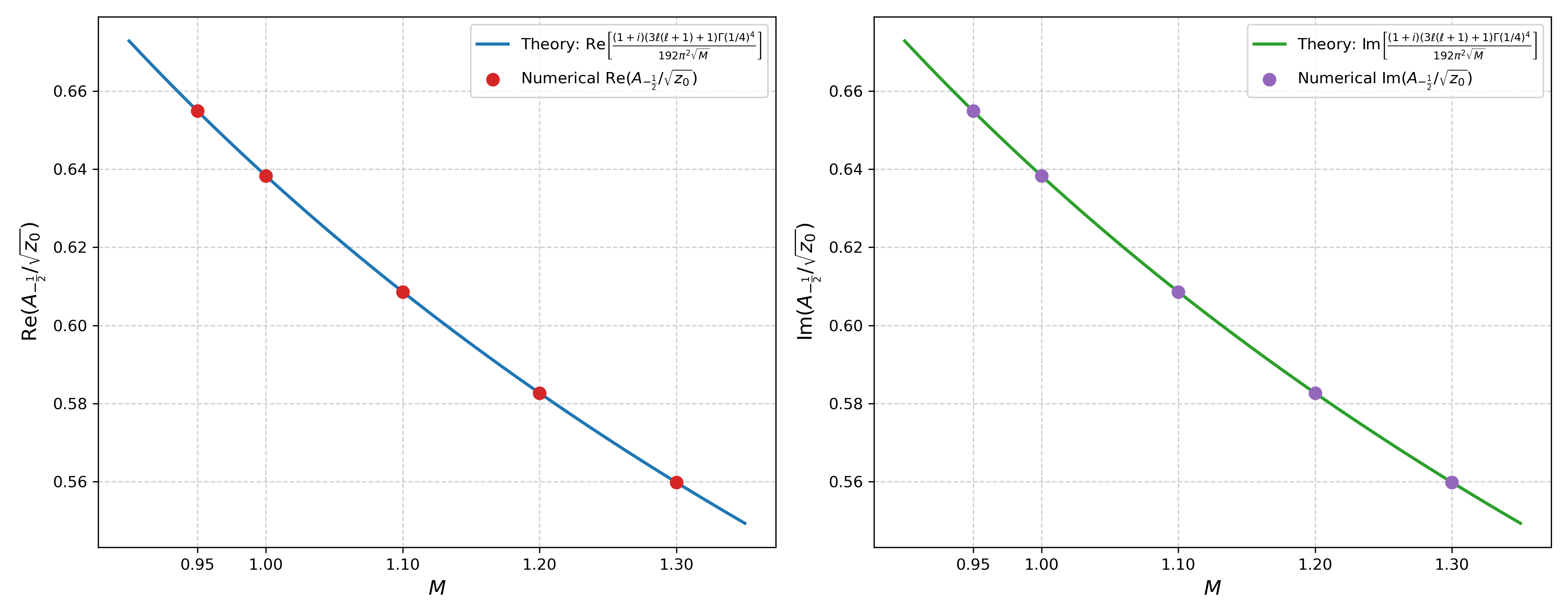}
        \caption{$A_{-\frac{1}{2}}/\sqrt{z_0}$ vs $M$}
        \label{fig:B2_M}
    \end{subfigure}
    \vspace{0.8em}
    \begin{subfigure}[b]{0.95\textwidth}
        \includegraphics[width=\textwidth]{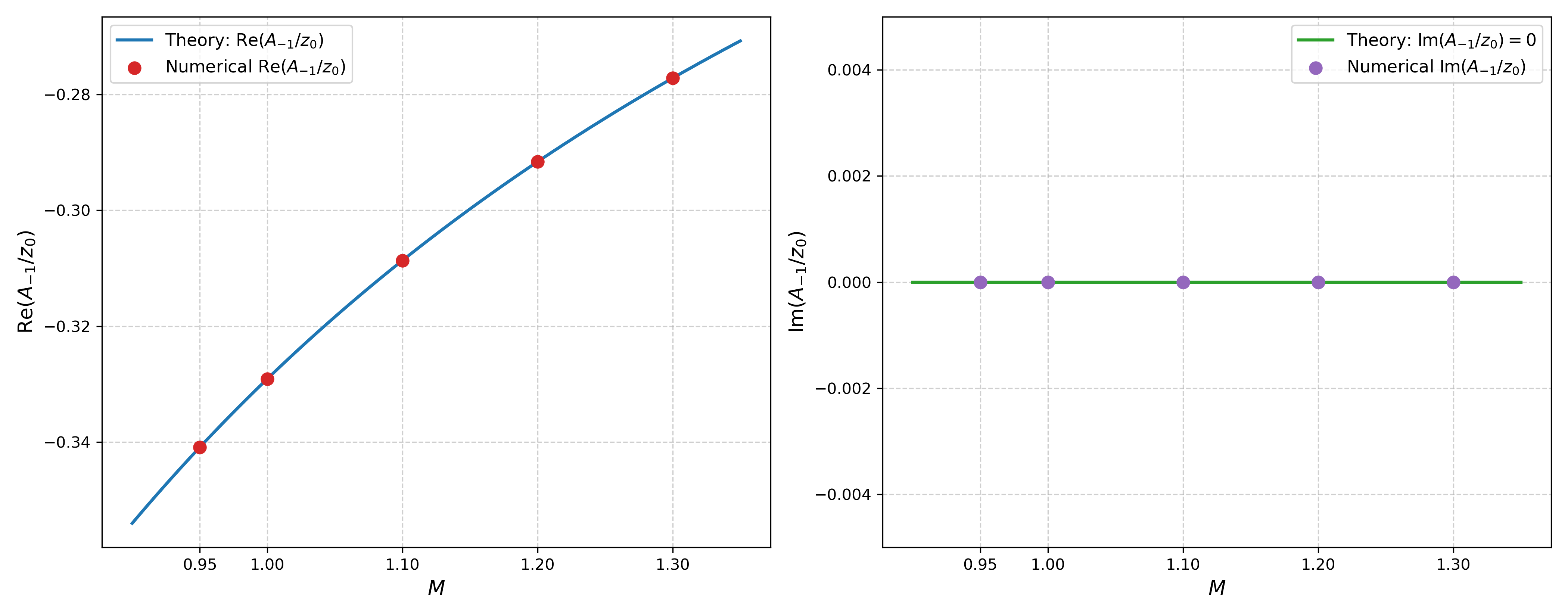}
        \caption{$A_{-1}/z_0$ vs $M$}
        \label{fig:B3_M}
    \end{subfigure}
    \caption{Normalized coefficients appearing in Eq.\eqref{Num-omega-exp} vs black hole mass $M \in [0.95, 1.30]$ for $\Lambda=0.06, \ell=1, m=0$. (a) $A_{-\frac{1}{2}}/\sqrt{z_0}$ decays as $M^{-1/2}$. (b) $A_{-1}/z_0$ is purely real }
    \label{fig:M_sweeps}
\end{figure}
In Figures~(\ref{fig:B2_M}) and (\ref{fig:B3_M}), the parameters $A_{-\frac{1}{2}}/\sqrt{z_0}$ and $A_{-1}/z_0$ are plotted against the black hole mass $M \in [0.95, 1.30]$ for fixed $\Lambda = 0.06, \ell = 1, m = 0$. Here, the numerical points (red circles for $\mathrm{Re}$, purple for $\mathrm{Im}$) lie exactly on the continuous analytic curves (blue and green) for all points. We can also see from Figure~(\ref{fig:B2_M}) for the parameter $A_{-\frac{1}{2}}/\sqrt{z_0}$ scales as $1/\sqrt{M}$. In Figure~\ref{fig:B3_M} we can see that $A_{-1}/z_0$ is purely real as predicted by Eq.~\eqref{QNM-exp}.
\begin{figure}[h!]
    \centering
    \begin{subfigure}[b]{0.95\textwidth}
        \includegraphics[width=\textwidth]{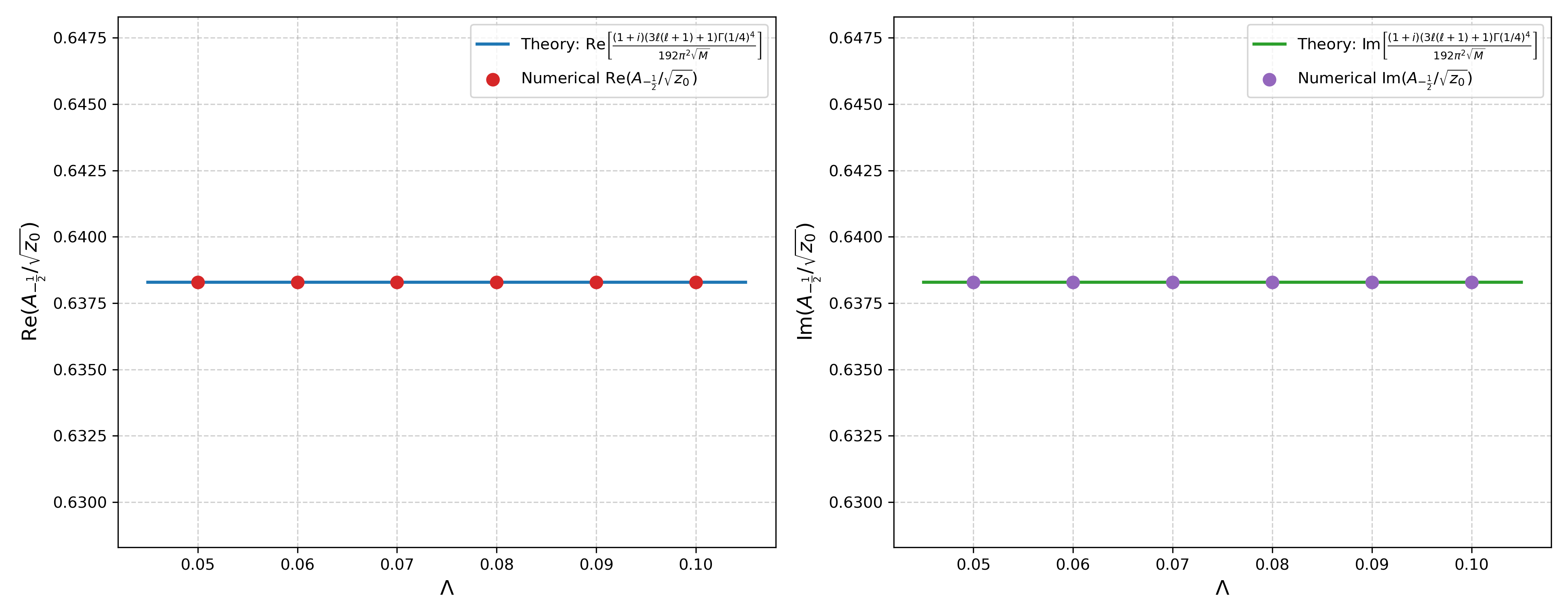}
        \caption{$A_{-\frac{1}{2}}/\sqrt{z_0}$ vs $\Lambda$}
        \label{fig:B2_lam}
    \end{subfigure}
    \vspace{0.8em}
    \begin{subfigure}[b]{0.95\textwidth}
        \includegraphics[width=\textwidth]{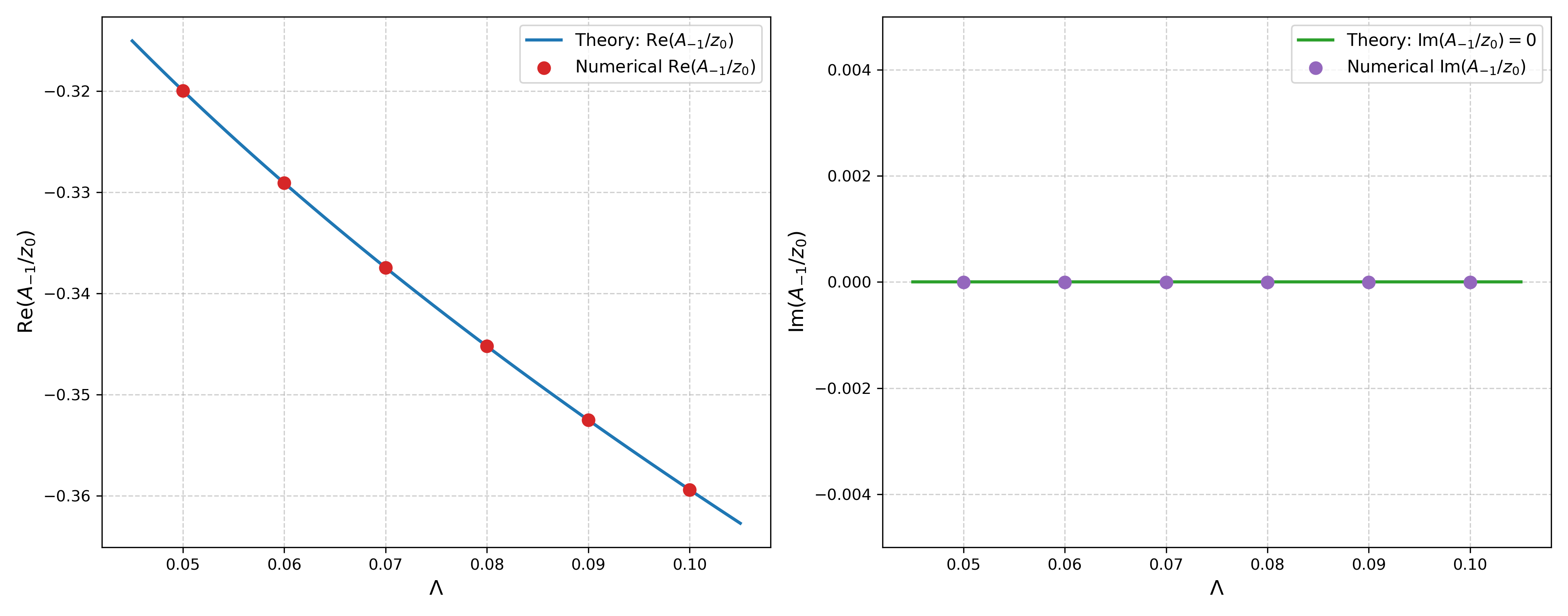}
        \caption{$A_{-1}/z_0$ vs $\Lambda$}
        \label{fig:B3_lam}
    \end{subfigure}
    \caption{Normalized coefficients vs cosmological constant $\Lambda \in [0.05, 0.10]$ for $M=1, \ell=1, m=0$. (a) $A_{-\frac{1}{2}}/\sqrt{z_0}$ is strictly flat and invariant under $\Lambda$. (b) $A_{-1}/z_0$ is purely real and varies smoothly due to the contraction of the cavity wall radius $r_\mathcal{O}(\Lambda) = (3M/\Lambda)^{1/3}$.}
    \label{fig:lam_sweeps}
\end{figure}
In Figure~\ref{fig:B2_lam}, we observe the following characteristic features: both the real and imaginary parts of the normalized coefficient $A_{-\frac{1}{2}}/\sqrt{z_0}$ lie on horizontal lines across the entire range $\Lambda \in [0.05, 0.10]$. This agrees with the fact that the plotted values of $A_{-\frac{1}{2}}/\sqrt{z_0}$ are not dependent on the $\Lambda$. And similarly the imaginary part of Figure~\ref{fig:B3_lam} is also zero because only the $A_{-1}/z_0$ is real and it depends on $\Lambda$.
\begin{figure}[H]
    \centering
    \begin{subfigure}[b]{0.48\textwidth}
        \includegraphics[width=\textwidth]{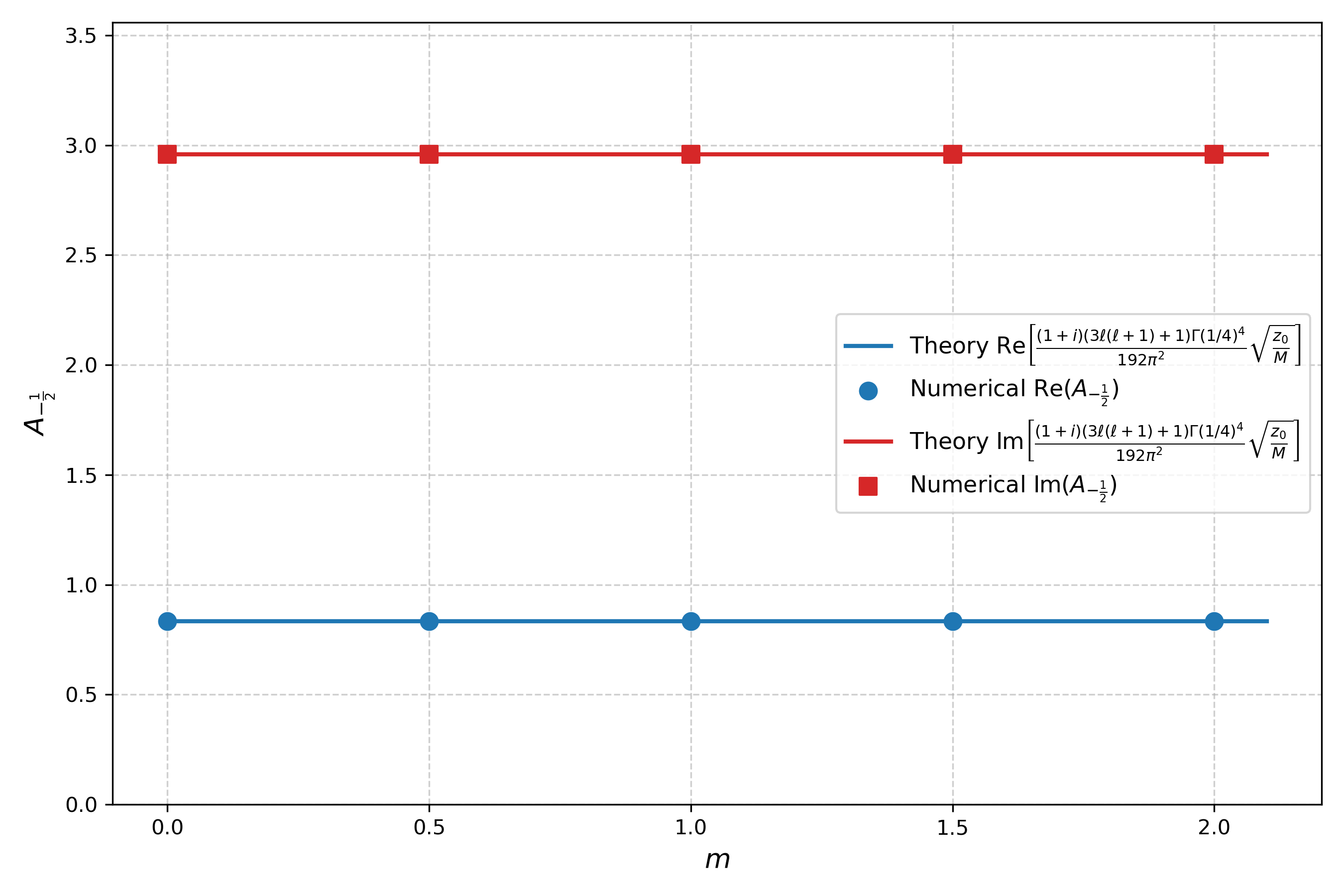}
        \caption{$A_{-\frac{1}{2}}$ vs $m$}
        \label{fig:B2_m}
    \end{subfigure}
    \hfill
    \begin{subfigure}[b]{0.48\textwidth}
        \includegraphics[width=\textwidth]{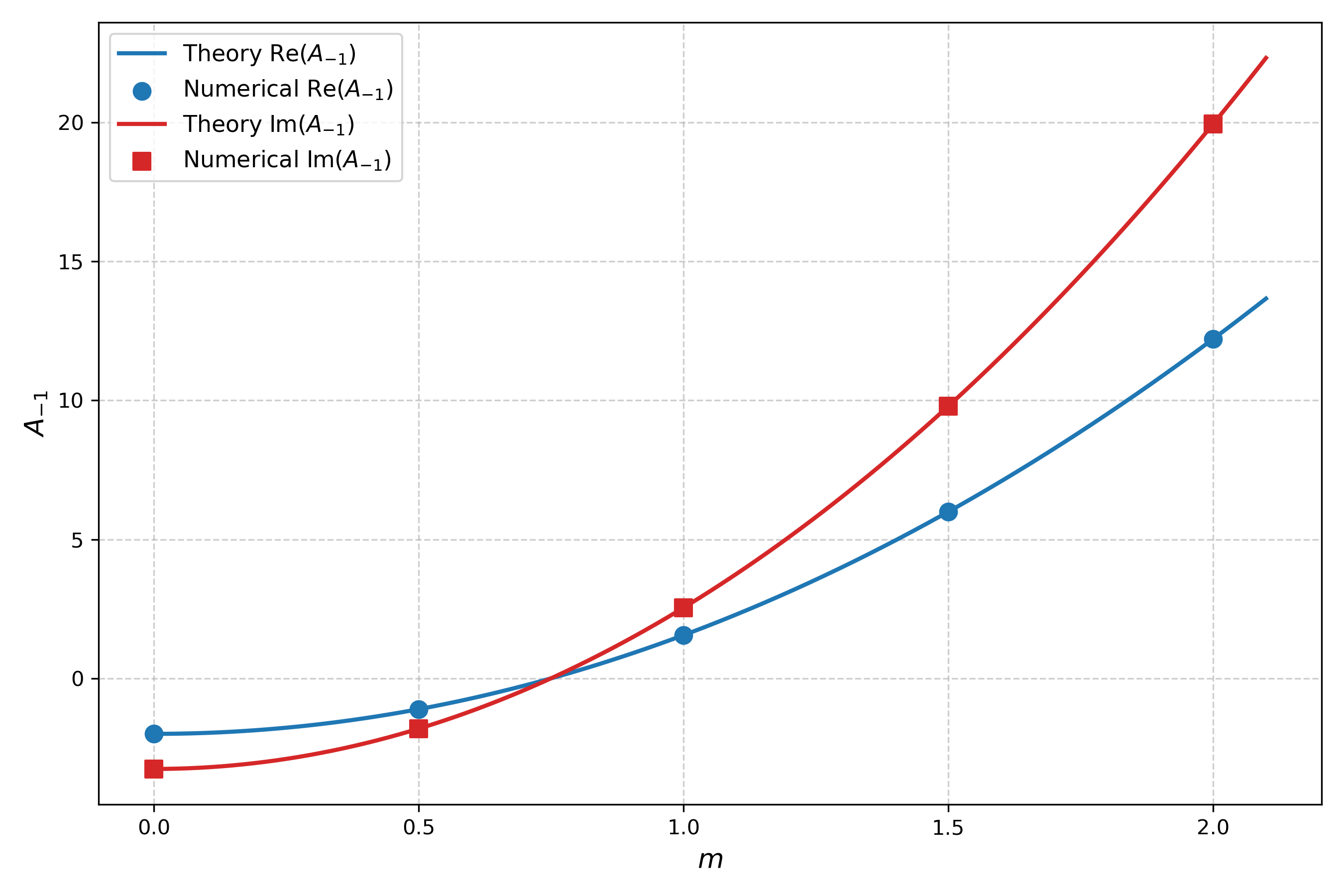}
        \caption{$A_{-1}$ vs $m$}
        \label{fig:B3_m}
    \end{subfigure}
    \caption{Coefficients $A_{-\frac{1}{2}}$ and $A_{-1}$ appearing in Eq.\eqref{Num-omega-exp} vs scalar field mass $m \in [0.0, 2.0]$ for $\ell=1, M=1, \Lambda=0.06$. (a) $A_{-\frac{1}{2}}$ is strictly constant with respect to $m$. (b) $A_{-1}$ displays an exact upward quadratic parabola $\propto m^2$.}
    \label{fig:m_sweeps}
\end{figure}

In Figure~(\ref{fig:B2_m}), the coefficient $A_{-\frac{1}{2}}$ is plotted against the scalar field mass $m \in [0.0, 2.0]$ for fixed $\ell = 1, M = 1, \Lambda = 0.06$. Both $\mathrm{Re}\left(A_{-\frac{1}{2}}\right)$ and $\mathrm{Im}\left(A_{-\frac{1}{2}}\right)$ are  horizontal lines implying the coefficient is  independent of scalar mass. In contrast, Figure~(\ref{fig:B3_m}) shows that $A_{-1}$ curves steeply upwards as $m$ increases. The formula in Eq.~\eqref{QNM-exp} contains the mass term in the form $m^2 r_\mathcal{O}$. We can see from the Figure~(\ref{fig:lam_sweeps}) and Figure~(\ref{fig:m_sweeps}) that contributions from $\Lambda$ (other than through $z_0$) and $m$ come only from $\mathcal{O}\left(\frac{1}{n}\right)$ terms. 
\section{Discussion}	
\label{Sec:discuss}
    We have studied large quasi-normal mode frequencies of a scalar probe field in Schwarzschild de-Sitter space by imposing Dirichlet boundary condition on a time-like surface placed at the static sphere radius. This boundary condition differs from that considered in \cite{Cardoso:2004up,Ghosh:2005aq}, because of which Stokes topology is different. Following the Stokes curves, we find connection coefficients relating the scalar wave function at black hole and cosmological horizons. The main result of this paper is the asymptotic expansion of the scalar quasi-normal modes at large overtones, given in Eq.\eqref{QNM-exp}. This has implications in the study of two-sided boundary correlator. At large imaginary energy, $\omega=-i E$, with $E\rightarrow \infty$, using Eq.\eqref{bound-cor} we can write 
	\begin{equation}
		\partial_{E}\log G_{12}^{\partial}\left(-i E\right) = -2\operatorname{Re}\left[\sum_{n=1}^{\infty}\frac{2E}{E^{2}+\omega_{n}^{2}}\right]~.
	\end{equation}
	Considering the leading asymptotic form from Eq.\eqref{QNM-exp}, $\omega_{n}=\frac{\pi n}{z_{0}}$ sum in the square bracket takes a closed form expression of $z_{0}\coth\left(Ez_{0}\right)-\frac{1}{E}$. Since in the sub-Nariai limit  $0<3GM\sqrt{\Lambda}<1$ so $\operatorname{Re}\left(z_{0}\right)>0$. This implies for large $E$ 
	\begin{equation}
		G_{12}^{\partial}\left(-i E\right) \sim \exp[-2E \operatorname{Re\left(z_{0}\right)}] \sim \exp\left[-E \operatorname{Re}\left(t_{\ast}\right)\right]~.
	\end{equation}
	Argument appearing in the exponential can be interpreted from WKB approximation at large imaginary frequency: $S\simeq\int_{0}^{z_{0}}\sqrt{E^{2}+V\left(z\right)}\mathrm{d}z \simeq Ez_{0}+\mathcal{O}\left(1\right)$, with $E^{2}\gg V\left(z\right)$ and we find it depends on the bouncing time. Subleading corrections to the  asymptotic  quasi-normal modes will give rise to  non-analytic terms in the two-sided correlator \cite{Hartnoll:2026vhu}. Going beyond the leading order, we will obtain $\log G^{\partial}_{12}(-iE)=\text{const}-E\,\operatorname{Re}t_*+p\log E-\sqrt2\,\alpha\gamma\,E^{-\frac{1}{2}}+\mathcal O(E^{-1})$, $p$ being a constant. Coefficient of $E^{-\frac{1}{2}}$ depends on local data near the black hole singularity, whereas the higher order terms will depend on the far-off region as well. We leave the analysis of this correlator in time domain for a future work. In this context, study of asymptotic QNM may be useful in understanding holographic correlators in dS spacetimes \cite{Aalsma:2022eru,Chapman:2022mqd, Goldar:2026bij}. 

    It will be interesting to extend the present analysis to gravitational perturbations. A preliminary analysis with the same Dirichlet boundary condition at the wall suggests that leading order high-overtone spacing of gravitational QNM is similar to that found in the scalar case reported here. The sub-leading constant offset term depends on parity of the gravitational perturbations: for polar sectors it matches with that of the scalar result, for axial sectors it differs by a sign in the real part. In both the cases, $\Lambda$ appears in the local near-singularity matching condition along Stokes curves at $n^{-\frac{3}{2}}$ order.

    \section*{Acknowledgements}
    We would like to thank Joydeep Chakravarty, Justin David, Leonard Schwarze and Aninda Sinha for helpful discussions.  We are particularly grateful to Aninda Sinha for several discussions on dS which led to the genesis of this project. We acknowledge use of AI agents such as ChatGPT, Claude and Gemini for verifying calculations and performing numerical checks. APS is supported by DST INSPIRE Faculty Fellowship (IFA22-PH 282). 
	
\appendix

\section{Stokes geometry for asymptotic QNM}
\label{App:Stokes}

\begin{figure}[H]
    \centering
    \includegraphics[width=0.95\linewidth]{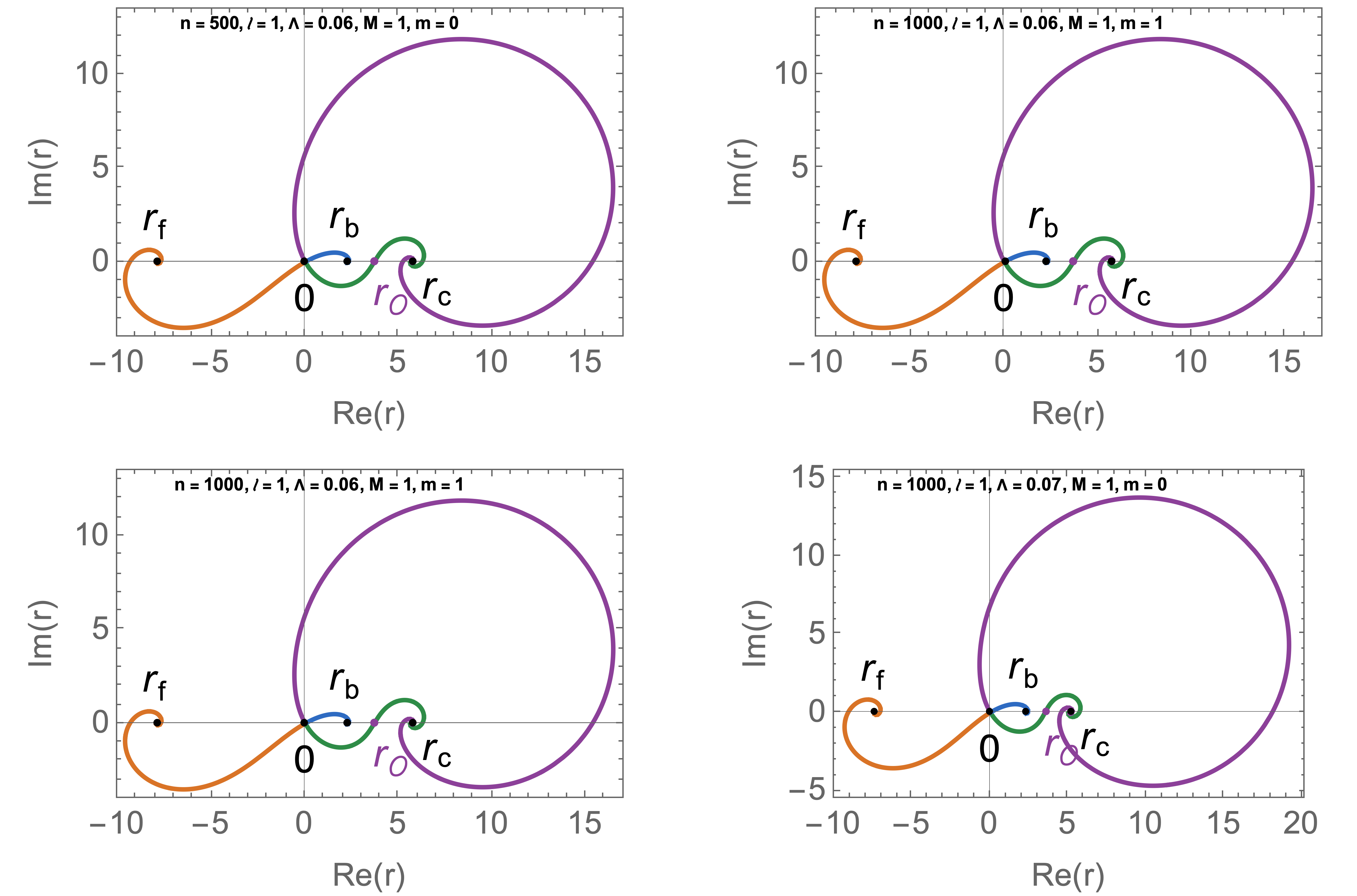}
    \caption{Stokes curves corresponding to the asymptotic quasi-normal modes plotted from different choices of large overtone number, angular momentum and mass of the scalar field.}
    \label{fig:Stokes-geom-app}
\end{figure}
Here we present some qualitative features of the Stokes geometry in the SdS with Dirichlet boundary wall at the static sphere. There are four Stokes curves emanating from $r=0$ as described by Eq.\eqref{stokes-angle}. On these curves $\operatorname{Im}\left(\omega_{n}\xi\right)=0$ for large integer values of $n$. As $\omega\xi=-\omega z_{0} \approx -\pi n$ at $r=r_{\mathcal{O}}$, therefore green Stokes curve almost passes through the wall at static sphere.  
\begin{table}[H]
\centering
\begin{tabular}{|c|c|}
\hline
Parameters & $\Delta r_{\mathcal{O}}$ \\
\hline
$n = 500$, $\ell = 1$, $\Lambda = 0.06$, $M = 1$, $m = 0$ & $7.60286 \times 10^{-4}$ \\
\hline
$n = 1000$, $\ell = 2$, $\Lambda = 0.06$, $M = 1$, $m = 0$ & $1.07838 \times 10^{-3}$ \\
\hline
$n = 1000$, $\ell = 1$, $\Lambda = 0.06$, $M = 1$, $m = 1$ & $7.78821\times 10^{-4}$ \\
\hline
$n = 1000$, $\ell = 1$, $\Lambda = 0.07$, $M = 1$, $m = 0$ & $6.26568 \times 10^{-4}$ \\
\hline
\end{tabular}
\caption{$\Delta r_{\mathcal{O}}$ is the absolute value of difference in the radial coordinates between the point where green Stokes curve intersects the real $r$ axis and $r_{\mathcal{O}}$. }
\end{table}

Let $u=\omega_{n}\left(z-z_{0}\right)\in\mathbb{R}$. Then from $\frac{\mathrm{d}z}{\mathrm{d}r}=-\frac{1}{f\left(r\right)}$ we can write,
\begin{equation}
    \mathrm{d}u=-\frac{\omega_{n}}{f\left(r\right)}\mathrm{d}r~.
\end{equation}
Near any horizon, we can approximate $f\left(r_{i}\right)\simeq f'\left(r_{i}\right)\left(r-r_{i}\right)$. This leads to 
\begin{equation}
    r-r_{i} \propto \exp\left[-\frac{f'\left(r_{i}\right)}{\omega_{n}}u\right]~.
\end{equation}
Taking $r-r_{i}=\rho e^{i\Delta\theta}$ and $\omega_{n}= \omega_{R}+ i \omega_{I}$, real and imaginary parts of $\omega_{n}$ respectively,  we obtain 
\begin{equation}
    \rho \propto \exp\left[-\frac{f'\left(r_{i}\right)\omega_{R}}{|\omega_{n}|^{2}}u\right]~, \qquad \Delta\theta = \frac{f'\left(r_{i}\right)\omega_{I}}{|\omega_{n}|^{2}}u~.
\end{equation}
Combining the above equations, we find
\begin{equation}
    \rho\propto \exp\left[-\frac{\omega_{R}}{\omega_{I}}\Delta\theta\right]~.
\end{equation}
Since, in the present case, quasi-normal modes have $\omega_{R}>0$ and $\omega_{I}<0$, therefore $\rho$ increases (decreases) with increasing (decreasing) value of $\Delta\theta$. This explains clockwise inward spiraling of the Stokes curves towards the horizons in Fig.(\ref{fig:Stokes-geom-app}). Each of the curves reaches the respective horizon only asymptotically and in the process winds around the horizon infinitely many times. This scenario is in contrast to the asymptotic quasi-normal modes with large damping term considered in \cite{Cardoso:2004up}. There $|\omega_{I}|\gg |\omega_{R}|$; therefore exponential term is a pure phase, and hence the Stokes lines close among themselves instead of terminating at the horizons.   

\section{Volterra iterated integral solution}
\label{App:Volterra}

We briefly review the construction of iterated Volterra solution using 	the standard method of variation of parameters \cite{2010arXiv1011.1775B}. Solutions to the homogeneous equation in Eq.\eqref{scalar-eom} are given by $p\left(u\right)$ and $q\left(u\right)$, defined in Eq.\eqref{homogeneousbasis}. Then the perturbed solution can be written as
\begin{equation}
	\psi\left(u\right)=c_{1}\left(u\right)p\left(u\right)+c_{2}\left(u\right)q\left(u\right)~,
\end{equation}
with the coefficients satisfying the constraint
\begin{equation}
	c'_{1}\left(u\right)p\left(u\right)+c'_{2}\left(u\right)q\left(u\right)=0~.
\end{equation}
Substituting $\psi\left(u\right)$ in Eq.\eqref{subleading-eom} gives
\begin{equation}
	\begin{pmatrix}
		p\left(u\right) & q\left(u\right)\\
		p'\left(u\right) & q'\left(u\right)
	\end{pmatrix}
	\begin{pmatrix}
		c'_{1}\left(u\right)\\ c'_{2}\left(u\right)
	\end{pmatrix}
	=-\frac{\alpha}{\sqrt{\omega}u^{\frac{3}{2}}}\begin{pmatrix}
		0 \\ p\left(u\right)c_{1}\left(u\right)+q\left(u\right)c_{2}\left(u\right)
	\end{pmatrix}~.
\end{equation}
Above equation implies the first order matrix differential equation,
\begin{equation}
	\begin{pmatrix}
		c'_{1}\left(u\right)\\ c'_{2}\left(u\right)
	\end{pmatrix} = \frac{\alpha}{\sqrt{\omega}u^{\frac{3}{2}}}
	\begin{pmatrix}
		p\left(u\right)q\left(u\right) & q\left(u\right)^{2}\\
		-p\left(u\right)^{2} & -p\left(u\right)q\left(u\right)
	\end{pmatrix}
	\begin{pmatrix}
		c_{1}\left(u\right)\\ c_{2}\left(u\right)
	\end{pmatrix}~.
\end{equation}
Let us define $\mathbf{c}\left(u\right)=\begin{pmatrix}
	c_{1}\left(u\right) \\ c_{2}\left(u\right)
\end{pmatrix}$. We introduce a transfer matrix, $\mathcal{U}\left(u;\alpha\right)$ which connects the coefficients at large $u$ to that at $u=0$,
\begin{equation}\label{coeff-transfer}
	\mathbf{c}\left(u\right)= \mathcal{U}\left(u;\alpha\right)\mathbf{c}\left(0\right)~, 
\end{equation}
with $\mathcal{U}\left(0;\alpha\right)=\mathbb{I}$, identity matrix. Then we get
\begin{equation}\label{evolutionEq}
	\frac{\mathrm{d}}{\mathrm{d}u}\mathcal{U}\left(u;\alpha\right)=\frac{\alpha}{\sqrt{\omega}}\mathbf{R}\left(u\right)\mathcal{U}\left(u,\alpha\right)~, \qquad \mathbf{R}\left(u\right)=\frac{1}{u^{\frac{3}{2}}}\begin{pmatrix}
		p\left(u\right)q\left(u\right) & q\left(u\right)^{2}\\
		-p\left(u\right)^{2} & -p\left(u\right)q\left(u\right)
	\end{pmatrix}~.
\end{equation}
This gives an evolution equation for $\mathcal{U}$. We can then write an integral equation
\begin{equation}
	\mathcal{U}\left(u;\alpha\right)=\mathcal{P}\exp\left[\frac{\alpha}{\sqrt{\omega}}\int_{0}^{u}\mathbf{R}\left(t\right)\mathrm{d}t\right],
\end{equation}
where $\mathcal{P}$ denotes $u$-ordering. Perturbatively $\mathcal{U}$ can be expressed as
\begin{equation}
	\begin{aligned}
		\mathcal{U}\left(u,\alpha\right)=& \mathbb{I}+\frac{\alpha}{\sqrt{\omega}}\int_{0}^{u}\mathrm{d}t\;\mathbf{R}\left(t\right) + \frac{\alpha^{2}}{\omega}\int_{0}^{u}\mathrm{d}u_{1}\; \mathbf{R}\left(u_{1}\right)\int_{0}^{u_{1}}\mathrm{d}u_{2}\;\mathbf{R}\left(u_{2}\right)\\
		& + \frac{\alpha^{3}}{\omega^{\frac{3}{2}}}\int_{0}^{u}\mathrm{d}u_{1}\; \mathbf{R}\left(u_{1}\right)\int_{0}^{u_{1}}\mathrm{d}u_{2}\;\mathbf{R}\left(u_{2}\right)\int_{0}^{u_{2}}\mathrm{d}u_{3}\;\mathbf{R}\left(u_{3}\right) + \cdots~,
	\end{aligned}
\end{equation}
with the ordering $u>u_{1}>u_{2}>u_{3}>\cdots>0$.

Let us introduce the following notations,
\begin{equation}\label{def-integrals}
	F_{1}\left(u\right) = \int_{0}^{u}\mathrm{d}t\; t^{-\frac{3}{2}}p\left(t\right)q\left(t\right)~, \qquad F_{2}\left(u\right)=\int_{0}^{u}\mathrm{d}t\; t^{-\frac{3}{2}}p\left(t\right)^{2}~, \qquad F_{3}\left(u\right) = \int_{0}^{u}\mathrm{d}t\; t^{-\frac{3}{2}}q\left(t\right)^{2}~.
\end{equation}
Using the Weber-Schafheitlin formula,
\begin{equation}
	\begin{aligned}
		\mathcal{I}_{\mu\nu}\left(\lambda\right) & := \int_{0}^{\infty}\mathrm{d}t\; t^{-\lambda}J_{\mu}\left(t\right)J_{\nu}\left(t\right) \\
		& = \frac{2^{-\lambda } \Gamma (\lambda ) \Gamma \left(\frac{1}{2} (-\lambda +\mu +\nu
			+1)\right)}{\Gamma \left(\frac{1}{2} \left(\lambda +\mu -\nu +1\right)\right) \Gamma
			\left(\frac{1}{2} \left(\lambda -\mu +\nu +1\right)\right) \Gamma \left(\frac{1}{2} \left(\lambda +\mu
			+\nu +1\right)\right)}
	\end{aligned}
\end{equation}
and $\partial_{\mu}J_{\mu}\left(t\right)\vert_{\mu=0}=\frac{\pi}{2}Y_{0}\left(t\right)$, we can evaluate the expressions in Eq.\eqref{def-integrals} in the large $u$ limit,
\begin{equation}
	\begin{aligned}
		F_{1}\left(\infty\right) & = -\frac{\sqrt{2} \pi ^{\frac{3}{2}} \Gamma \left(\frac{5}{4}\right)}{\Gamma
			\left(\frac{3}{4}\right)^3} =-\gamma~,\\
		F_{2}\left(\infty\right) & = \frac{\sqrt{2} \pi ^{\frac{3}{2}} \Gamma \left(\frac{5}{4}\right)}{\Gamma
			\left(\frac{3}{4}\right)^3} = \gamma~,\\
		F_{3}\left(\infty\right) & = \frac{3 \sqrt{2} \pi ^{3/2} \Gamma \left(\frac{5}{4}\right)}{\Gamma
			\left(\frac{3}{4}\right)^3} = 3\gamma~.
	\end{aligned}
\end{equation}
From the above relations,  we obtain 
\begin{equation}\label{H1-mat}
	H_{1}:= \int_{0}^{\infty}\mathrm{d}t\; \mathbf{R}\left(t\right)=
	\gamma\begin{pmatrix}
		-1 & 3\\ -1 & 1
	\end{pmatrix}~.
\end{equation}
At the next order in $\omega^{-\frac{1}{2}}$, we have 
\begin{equation}\label{H2-eq}
	\begin{aligned}
		H_{2} &:= \int_{0}^{u}\mathrm{d}u_{1}\; \mathbf{R}\left(u_{1}\right)\int_{0}^{u_{1}}\mathrm{d}u_{2}\;\mathbf{R}\left(u_{2}\right) \\
		& = \int_{0}^{\infty}\frac{\mathrm{d}u}{u^{\frac{3}{2}}} 
		\begin{pmatrix}
			q\left(u\right) \left(F_{1}\left(u\right) p\left(u\right)-F_{2}\left(u\right) q\left(u\right)\right) & q\left(u\right) \left(F_{3}\left(u\right) p\left(u\right)-F_{1}\left(u\right) q\left(u\right)\right) \\
			p\left(u\right) \left(-F_{1}\left(u\right) p\left(u\right)+F_{2}\left(u\right) q\left(u\right)\right) & p\left(u\right) \left(F_{1}\left(u\right) q\left(u\right)-F_{3}\left(u\right) p\left(u\right)\right)
		\end{pmatrix}~.
	\end{aligned}
\end{equation}
To evaluate the above matrix elements, we define
\begin{equation}
	\begin{aligned}
		\mathcal{W}_{\alpha\beta;\gamma\delta} & :=\int_{0}^{\infty}\mathrm{d}u_{1}\;u_{1}^{-\frac{1}{2}}J_{\alpha}\left(u_{1}\right)J_{\beta}\left(u_{1}\right)\int_{0}^{u_{1}}\mathrm{d}u_{2}\;u_{2}^{-\frac{1}{2}}J_{\gamma}\left(u_{2}\right)J_{\delta}\left(u_{2}\right)\\
		& = \int_{0}^{\infty}\mathrm{d}u_{1}\;u_{1}^{-\frac{1}{2}}J_{\alpha}\left(u_{1}\right)J_{\beta}\left(u_{1}\right)\int_{0}^{\infty}\mathrm{d}u_{2}\;u_{2}^{-\frac{1}{2}}J_{\gamma}\left(u_{2}\right)J_{\delta}\left(u_{2}\right)\Theta\left(u_{1}-u_{2}\right)~,
	\end{aligned}
\end{equation}
where $\Theta$ is the Heaviside step function and has an inverse Mellin representation, 
\begin{equation}
	\Theta\left(u_{1}-u_{2}\right) = \frac{1}{2\pi i}\int_{c-i\infty}^{c+i\infty}\frac{\mathrm{d}s}{s}\left(\frac{u_{1}}{u_{2}}\right)^{s}~, \qquad 0<c<\frac{1}{2}~.
\end{equation}
Then we can write
\begin{equation}
	\mathcal{W}_{\alpha\beta;\gamma\delta} = \frac{1}{2\pi i}\int_{c-i\infty}^{c+i\infty}\frac{\mathrm{d}s}{s}\mathcal{I}_{\alpha\beta}\left(\frac{1}{2}-s\right)\mathcal{I}_{\gamma\delta}\left(\frac{1}{2}+s\right)~.
\end{equation}
Using the above identity and taking large $u$ limit, Eq.\eqref{H2-eq} reduces to 
\begin{equation}\label{H2-derivativeexp}
	H_{2} = \begin{pmatrix}
		\partial_{\beta}\partial_{\delta}-\partial_{\alpha}\partial_{\beta} & \frac{2}{\pi}\left(\partial_{\beta}\partial_{\gamma}\partial_{\delta}-\partial_{\alpha}\partial_{\beta}\partial_{\delta}\right) \\
		\frac{\pi}{2}\left(-\partial_{\delta}+\partial_{\beta}\right) & \left(\partial_{\beta}\partial_{\delta}-\partial_{\gamma}\partial_{\delta}\right)
	\end{pmatrix}\mathcal{W}_{\alpha\beta;\gamma\delta}\vert_{\{\alpha,\beta\gamma,\delta\}\rightarrow 0}~.
\end{equation}
For later use, we note the following identities
\begin{equation}
	\begin{aligned}
		\partial_{\alpha}\mathcal{I}_{\alpha 0}\left(\lambda\right)\rvert_{\alpha=0} &= -\frac{\pi}{2}\tan\left(\frac{\pi\lambda}{2}\right)\mathcal{I}_{00}\left(\lambda\right)~, \\ \partial_{\alpha}\partial_{\beta}\mathcal{I}_{\alpha\beta}\left(\lambda\right)\rvert_{\{\alpha,\beta\}= 0} &= \frac{\pi^{2}}{4}\left(1+2\tan^{2}\left(\frac{\pi\lambda}{2}\right)\right)\mathcal{I}_{00}\left(\lambda\right)~.
	\end{aligned}
\end{equation}
Using these, Eq.\eqref{H2-derivativeexp} leads to the following expression,
\begin{equation}\label{H2-mat}
	\begin{aligned}
		H_{2} & = \frac{1}{2\pi i}\int_{c-i\infty}^{c+i\infty}\frac{\mathrm{d}s}{s}
		\begin{pmatrix}
			-\frac{\pi  \Gamma
				\left(\frac{s}{2}+\frac{1}{4}\right)^4}{8 \Gamma
				\left(\frac{s}{2}+\frac{3}{4}\right)^4} & -\frac{\pi ^3 \tan (\pi  s) \sec (\pi  s) \Gamma
				\left(\frac{1}{4}-\frac{s}{2}\right) \Gamma
				\left(\frac{s}{2}+\frac{1}{4}\right)}{4 \Gamma \left(\frac{3}{4}-\frac{s}{2}\right)^3
				\Gamma \left(\frac{s}{2}+\frac{3}{4}\right)^3}\\
			\frac{\pi ^3 \tan (\pi  s) \sec (\pi  s) \Gamma
				\left(\frac{1}{4}-\frac{s}{2}\right) \Gamma
				\left(\frac{s}{2}+\frac{1}{4}\right)}{4 \Gamma
				\left(\frac{3}{4}-\frac{s}{2}\right)^3 \Gamma
				\left(\frac{s}{2}+\frac{3}{4}\right)^3} & -\frac{\pi  \Gamma
				\left(\frac{1}{4}-\frac{s}{2}\right)^4}{8 \Gamma
				\left(\frac{3}{4}-\frac{s}{2}\right)^4}
		\end{pmatrix} \\
		& = \gamma^{2}\begin{pmatrix}
			0 & -\frac{2}{3}\\
			\frac{2}{3} & -2
		\end{pmatrix}~.
	\end{aligned}
\end{equation}
In the above matrix, $\left(1,1\right)$ element has singularities at $s=0$, $s=-\frac{1}{2}-2n$, and $\left(2,2\right)$ element has poles at $s=0$, $s=\frac{1}{2}+2n$ for $n=0,1,2,\cdots$. For the first integral we can close the contour on the right half plane, and since there are no singularities there, it gives zero. For the second integral if we close the contour on the left half plane, then only pole at $s=0$ is encountered. $\left(1,2\right)$
 and $\left(2,1\right)$ elements differ by a sign. We choose the contour at infinity on the right side, which encloses poles at $s=\frac{1}{2}+2n$, $n=0,1,2,\cdots$.
 
Asymptotic form of the wave function near black hole horizon fixes 
\begin{equation}
    \begin{pmatrix}
			c_{1}\left(\infty\right)\\ c_{2}\left(\infty\right)
		\end{pmatrix} 
		 = \mathcal{N}_{b}\begin{pmatrix}
			1 \\ i
		\end{pmatrix}~,
\end{equation}
Using this we can find
\begin{equation}
		\begin{pmatrix}
			c'_{1}\left(\infty\right) \\ c'_{2}\left(\infty\right)
		\end{pmatrix}
		 =\mathcal{N}_{b}
		\begin{pmatrix}
			-3 i-2\left(1+ i\right) \gamma  g-\frac{4 \gamma ^2
				g^2}{3} + \mathcal{O}\left(g^{3}\right) \\
			1 +2\left(1- i\right) \gamma  g-\frac{4}{3} i \gamma ^2 g^2 + \mathcal{O}\left(g^{3}\right)
		\end{pmatrix}~, \quad g=\frac{\alpha}{\sqrt{\omega}}~.
\end{equation}
On the Stokes curve marked in green in Fig.(\ref{Fig:StokesGeom}), oscillatory solution near the singularity is given by 
\begin{equation}
	\psi \sim \sqrt{\frac{\pi s}{2}}\left[c'_{1}\left(s\right)J_{0}\left(s\right)+c'_{2}\left(s\right)Y_{0}\left(s\right)\right]~, \quad s=-\omega\xi>0~.
\end{equation}
Taking the large $s$ limit and matching with the wave function at the cosmological horizon gives 
\begin{equation}\label{D-coeffinf}
	\begin{aligned}
		D_{1} & \propto  e^{-i \omega z_{0}} + \mathcal{O}\left(g^{3}\right)~,\\
		D_{2} & \propto -2\left(i+\left(1+i\right)\gamma g+\frac{2}{3}\gamma^{2}g^{2}\right)e^{i\omega z_{0}} + \mathcal{O}\left(g^{3}\right)~.
	\end{aligned}
\end{equation}
Plugging the expressions of $D_{1}$ and $D_{2}$ and imposing the Dirichlet boundary condition at $z=0$ gives Eq.\eqref{potentialcorr}.

\bibliographystyle{JHEP}
\bibliography{Ref_SdS}
\end{document}